\documentclass[10pt]{article}
\usepackage[left=2.2cm,right=2.2cm,top=2.5cm,bottom=2.5cm]{geometry}

\usepackage{booktabs}
\usepackage{amsmath}
\usepackage{amsfonts}
\usepackage{amsthm}

\usepackage{url}
\usepackage{subfigure}
\usepackage{graphicx}
\usepackage[linesnumbered,ruled,vlined]{algorithm2e}
\usepackage{algpseudocode}
\usepackage[authoryear,square]{natbib}
\usepackage{wrapfig}
\usepackage{color}
\usepackage{epstopdf}
\usepackage[titletoc,toc,title]{appendix}
\usepackage{enumitem}
\usepackage{authblk}
\usepackage{epstopdf}
\usepackage{epsfig}
\usepackage{multicol}
\usepackage{multirow}
\usepackage{subfigure}
\usepackage{hyperref}

\renewcommand{\cite}{\citep}

\newtheorem{theorem}{Theorem}
\newtheorem{lemma}[theorem]{Lemma}

\newtheorem{remark}{Remark}

\begin{document}






%

\title{Searching for a Single Community in a Graph}

\author{Avik Ray \thanks{avik@utexas.edu}}
\author{Sujay Sanghavi \thanks{sanghavi@mail.utexas.edu}}
\author{Sanjay Shakkottai \thanks{shakkott@austin.utexas.edu}}
\affil{{Department of Electrical and Computer Engineering} \authorcr
{The University of Texas at Austin} \authorcr
{Austin, TX 78712}}

\date{$9^{th}$ April $2018$}

\maketitle
\begin{abstract}
In standard graph clustering/community detection, one is interested in partitioning the graph into more densely connected subsets of nodes. In contrast, the {\em search} problem of this paper aims to only find the nodes in a {\em single} such community, the target, out of the many communities that may exist. To do so , we are given suitable side information about the target; for example, a very small number of nodes from the target are labeled as such.

We consider a general yet simple notion of side information: all nodes are assumed to have random weights, with nodes in the target having higher weights on average. Given these weights and the graph, we develop a variant of the method of moments that identifies nodes in the target more reliably, and with lower computation, than generic community detection methods that do not use side information and partition the entire graph. Our empirical results show significant gains in runtime, and also gains in accuracy over other graph clustering algorithms.

\end{abstract}

%
%


\section{Introduction} \label{sec:intro}

Community detection, or graph clustering, is the classic problem of finding subsets of nodes such that each subset has higher connectivity within itself, as compared to the average connectivity of the graph as a whole. Typically, when graphs represent similarity or affinity relationships between nodes, these subsets represent communities of similar nodes. Also typically, this problem has primarily been considered in the unsupervised setting, where the only input is the graph itself and the objective is to partition all or most of the nodes. 

In this paper we look at a different, but related, community detection task, which we will refer to as the {\em search problem}. Our objective is to use the graph to find a single community of nodes -- which we will call the  {\em target community} -- for which we have been given some relevant but quite noisy side information. We would like to do so more reliably, and with lower computation, than existing methods that do not use side information.

Our motivations are two-fold: {\em (i)} it is often the case that the network analyst is looking for nodes with a-priori specified characteristics, and {\em (ii)} it is rare that we are faced with a ``pure'' graph analysis problem; typically there is extra non-graphical side information that, if used properly, could make the inference task easier.

As an example setting, consider the case where we have some nodes from the target community explicitly marked as such, and our task is to recover the remaining nodes. This is a situation that frequently arises in military/intelligence settings, and also in analysis of regular consumer social networks, internet/web graphs etc. In military intelligence it can be useful to recover a single community which a known suspect is part of. Besides explicit node labels, side information could also come from meta-information one may have about the nodes; e.g. from text analysis if the graph is a web graph, or from browse/activity history of users in a social network. In recommendation system or targeted advertising it is useful to learn a community of users with a specific interest (e.g. sports) using the knowledge of how users interact with relevant contents (e.g. sports news and images). Our aim is to find a principled way to use such side information {\em and} the graph itself.

{\bf Our contributions} are as follows.

\begin{itemize}
\item[(i)] We develop a simple yet generic framework for how side information is to be specified: each node is given a (possibly random) weight, with nodes in the target community having higher weight on average than nodes not in the target -- we call these {\em biased weights}. This setting would thus split an overall data + graph analysis objective into two: the analyst needs to devise a (application-dependent) procedure to convert her side information into biased node weights; these are then used by our algorithm.

\item[(ii)] Given such biased weights, we develop a new spectral-like algorithm -- specifically, a variant of the $2^{nd}$ order method of moments -- to find the nodes in the target community. We call this {\em Community Search} below. In the following, we first provide the basic intuition behind it by considering the case where we have access to the population statistics of a graph coming from a stochastic block model, and then formally describe the algorithm.

\item[(iii)] Our main results characterize the effectiveness of this algorithm in finding the target community; we study this in the standard stochastic block model setting with many communities. Analytically, we show that it matches (potentially up to log factors) the analytical guarantees of the state of the art unsupervised community detection methods; empirically, we show that the method outperforms these methods even with very noisy side information (e.g. very small number of labeled nodes), and has significantly lower computational complexity.

\item[(iv)] We also specialize our results to the case where the side information is in the form of a small number of labeled nodes; for this case we show how one can effectively convert this to node weights, even for sparse graphs. Our experiments on a real world network further corroborate the practical applicability of this method. 

\end{itemize}

\subsection{Related work}


While no other work has considered the problem of searching a single community in a graph, there has been a lot of research in three closely related fields; that of unsupervised and semi-supervised graph clustering, method of moments, and learning with side information. Each of these threads have a rich history -- here we cover the ones most relevant to this paper.

{\bf Unsupervised graph clustering:} Graph clustering or community detection has been widely studied mainly in the unsupervised setting where nodes do not have any associated labels. There is a vast literature of graph clustering algorithms both in the setting where clusters are non-overlapping \cite{Fort10} and overlapping \cite{XieKelSzy13}. The most widely studied generative model for non-overlapping clusters in a graph is the planted partition or stochastic block model \cite{ConKar:01}. Assuming this model many algorithms have been proposed which provide statistical guarantees of recovery of all hidden clusters. These algorithms can be broadly divided into three categories (i) spectral clustering \cite{McShe01,NgJorWei02,RoheChatYu11,ChaChuTsi12,AminiChenBickLevina:13,YunPro:14} (ii) convex optimization \cite{ChenSanXu12,AilChenXu13,AbbeBanHall:14} and more recently (iii) tensor decomposition \cite{AndGeHsuKak13,HuaNirHakAna13}. 

{\bf Semi-supervised graph clustering:} The graph clustering problem has also been explored in a semi-supervised settings, where some of the nodes and/or edges are explicitly labeled. Many optimization and kernel based algorithms have been proposed \cite{Zhu05,KulisBasuDhiMoo09} to solve this problem. The popular label propagation based clustering algorithms \cite{ZhouBouLalWesSch04,FujiIri14} are also essentially semi-supervised graph clustering algorithms with labeled nodes. Another related line of work also studies the graph clustering problem where the nodes have additional node features/attributes \cite{McAuleyLes:12,XuKeWang12,YangMcLes13,ZhangLevinaZhu:15}. More recently, local graph clustering algorithms based on message passing has also been studied in the semi-supervised setting \cite{CalLelMio:16, MosXu:16, CaiLiaRakh:16, KadAvrCotSun:18}.   

{\bf Method of Moments:} This is a classical parameter estimation technique, where the parameters to be estimated are described in terms of the moments from the true distribution. Empirical moments are now used to replace the true moments, leading to parameter estimates \cite{BowmanShenton:04}.  There has been much recent interest in these methods for many statistical learning problems. These include learning Gaussian mixture models \cite{HsuKakade13,AnaGeHsuKakTel14}, LDA topic models \cite{AnaLiuHsuFosKak12}, hidden Markov models \cite{Chang96} etc.

{\bf Others:} There is a broader machine learning literature that incorporates the availability of extra side information into existing models and algorithms. In the context of LDA topic models, side information maybe available in the form of extra response variables for each document \cite{BleiMcAuliffe08}, or additional text review information of products \cite{LuZhai08}. In collaborative filtering, side information can be of the form of item or user graph \cite{RaoYuRavDhi15}. 

In this paper we consider the community search problem with side information either in the form of biased node weights or a small set of labeled nodes.

\section{Settings and Algorithm} \label{sec:algo}

{\bf Stochastic Block Model:} Consider a graph $G=(V,E)$ with $n$ nodes and $k$ non-overlapping communities that partition the vertex set as $V=\cup_{i=1}^k V_i$. Let $\alpha_i=|V_i|/n$ be the fraction of nodes in the $i$-th community. In a stochastic block model the edge set $E$ is generated as follows. Let $0<q<p<1.$ Then for any two nodes in the same community $r,s \in V_i$ we have $P((r,s) \in E)=p,$ and when $r,s$ are in different community then $P((r,s) \in E)=q.$ We define this as the $(n,k,p,q)$ stochastic block model.

{\bf Target community and side information:} In the search problem we are interested in the recovery of {\em one} target community, in this paper, without loss of generality, consider $V_1$ to be this target community. We are also provided with some side information on this target community $V_1.$ The side information is in the form of {\em biased node weights.} Suppose for each node $j \in V$ we are given a biased weight $w_j>0.$ These weights are generated by a random process satisfying the condition that for any node $j \in V$ we have $E[w_j|j \in V_1]>E[w_j | j \in V_i]$ for all $i \neq 1.$

These biased weights may be computed using a set of {\em labeled nodes} from the target community $\mathcal{L} \subset V_1$ (see Section~\ref{sec:recovery_labeled_node}). These weights can also arise from other available sources of side information. For example consider a social network graph where the target community consists of users who are sports enthusiasts. Then we can observe the amount of interaction (e.g. ``likes'' and ``shares'' in Facebook) of the users with known sports related contents. Since users in a sports community are more likely to interact with such contents, these will have the above biased node weight property.
The main goal is to solve this search problem faster than the time required to recover all $k$ communities, and without any loss in estimation accuracy. 

\subsection{Algorithm}

In this section we describe our main algorithm called {\bf Community Search}. Let $X$ be the adjacency matrix of the graph $G.$ Also define community membership vectors $\mu_1, \hdots , \mu_k$ where $\mu_i \in \mathbb{R}^n,$ as follows. Let $\mu_{j,i}$ be the $j-$th coordinate of vector $\mu_i.$ Then,

\begin{equation*}
\mu_{j,i} =
\left\{
	\begin{array}{ll}
		p  & \mbox{if } j \in V_i \\
		q & \mbox{otherwise}
	\end{array}
\right.
\end{equation*}

Note that these $\mu_i$-s are linearly independent and the community memberships of the nodes can be obtained from these membership vectors via thresholding. The main purpose of our algorithm is to estimate the membership vector of the first community $\mu_1$ (which can then be used to recover nodes in $V_1$). 

{\bf Intuition behind our method:} To understand the core of our technique, let us suppose here -- just for intuition -- that we actually had access to the ``average" adjacency matrix $E[X]$ (recall that $X$ is the actual adjacency matrix of the stochastic block model), and let $E[X_j]$ be the average of the $j^{th}$ column. Then it is easy to see that $E[X_j] = \mu_{c_j}$, where $c_j \in 1,\ldots,k$ is the community that node $j$ belongs to. This means that the following holds for the matrix $A$ defined below:
\[
A ~ := ~ \frac{1}{n} \sum_{j=1}^n E[X_j] E[X_j]^T ~ = ~ \sum_{i=1}^k \alpha_i \mu_i \mu_i^T
\]
Similarly, let us now also suppose that we see the ``average" node weights $\bar{w}_j = E[w_j]$ for every node $j$. Then, the following holds for the matrix $B$ defined below:
\[
B ~ := ~ \frac{1}{n} \sum_{j=1}^n \bar{w}_j E[X_j] E[X_j]^T ~ = ~ \sum_{i=1}^k \alpha_i \omega_i \mu_i \mu_i^T
\]
where in the above, for each cluster $i$, we have defined $\omega_i$ be the averaged weights of all nodes in that cluster. By the bias condition, we have that $\omega_1 > \omega_i$ for all $i\neq 1$. 

Note that both the $A$ and $B$ as defined above are symmetric positive definite rank-$k$ matrices, with the column space of each spanned by the $\mu_i$'s. However note also that our desired vector $\mu_1$ may not be an eigenvector of either $A$ or $B$; indeed if the target community $V_1$ is small, it may be quite far from the leading eigenvector of either matrix. 

The main idea is that we can still recover $\mu_1$ by {\em ``whitening" $B$ using $A$}, a process we describe in the proto-algorithm below. The description also provides the (simple) reason why it works -- in this idealized case where average $X$ and $w$ are available. \\

\noindent {\bf Proto-algorithm (and explanation):}
\begin{enumerate}
\item Compute matrices $A$ and $B$ as described above,
\item Perform rank-$k$ svd of $A$ as $A=UDU^T,$ and let $W :=U D^{-1/2}.$ Also note that,
$$
W^T A W ~ = ~ I_k ~ = ~ \sum_{i=1}^k \tilde{\mu}_i \tilde{\mu}_i^T
$$
where we define $\tilde{\mu}_i :=\sqrt{\alpha_i} W^T \mu_i$. Now, we see that the addition of $k$ terms of the type $\tilde{\mu}_i \tilde{\mu}_i^T$ results in $I_k$; this can {\em only} happen if the corresponding $\tilde{\mu}_i$ are {\em orthonormal} vectors in $\mathbb{R}^k$. The vectors $\tilde{\mu}$ are thus ``whitened" versions of the original $\mu$ vectors.
\item Next we compute the following matrix.
$$
R ~ := ~ W^T B W ~ = ~ \sum_{i=1}^k \omega_i \tilde{\mu}_i \tilde{\mu}_i^T
$$ 
Now, since $\tilde{\mu}_i$ are orthonormal, the above equation represents an eigenvalue  decomposition of the $k \times k$ size matrix $R$, with eigenvectors $\tilde{\mu}_i$ and corresponding eigenvalues $\omega_i.$ Thus, $\tilde{\mu}_1$ -- the whitened vector corresponding to the target community -- is now the {\em leading eigenvector of $R$}, because $\omega_1 > \omega_i$.
\item Find $\tilde{\mu}_1$ by setting it to be the leading eigenvector of $R$. Finally we can recover $\mu_1$ from $\tilde{\mu}_1$ in two steps. First compute $z := U D^{1/2} \tilde{\mu}_1 = \sqrt{\alpha_1} \mu_1.$ Next compute vector 
$$
m_1 ~ := ~ \frac{1}{n} \sum_{j=1}^n E[X_j] ~ = ~ \sum_{i=1}^k \alpha_i \mu_i 
$$
We can recover $\sqrt{\alpha_1} = \tilde{\mu}_1^T W^T m_1.$ Then simply divide the $z$ defined above by this to find $\mu_1.$ 
\end{enumerate}  

{\bf An issue:} Although simple, it is not straight forward to convert this intuition to an  algorithm because due to inter dependencies it becomes hard to estimate these $A$ and $B$ matrices. In particular note that in the actual problem we are given the adjacency matrix $X$, and a natural impulse is to approximate A using the matrix 
\[
\frac{1}{n} \sum_{j=1}^n X_j X_j^T
\]
Unfortunately, this is a good approximation to $E[XX^T]$, but $E[XX^T] \neq E[X] E[X]^T$ -- and we require the latter.
However we can get around these dependencies by first partitioning the graph. This is outlined in Algorithm \ref{alg:comm_search_white} below. 

For any two subsets $P,Q \subset [n]$ let $X_{P,Q}$ denote the submatrix of $X$ corresponding to the rows and columns in set $P$ and $Q$ respectively. The input parameters to Algorithm \ref{alg:comm_search_white} are the adjacency matrix $X,$ number of communities $k,$ the set of biased node weights $(w_1 , \hdots , w_n),$ and a threshold $\tau.$ The output is the community estimate $\hat{V}_1.$ 

\IncMargin{1em}
\begin{algorithm}[ht]
\SetAlgoLined
\KwIn{Adjacency matrix $X,$ $k$, biased weights $(w_1,\hdots,w_n),$ threshold $\tau$} 
\KwOut{$\hat{V}_1$}
Partition nodes into four sets $P_1,P_2,P_3,P_4$ at random \;
Compute matrices $\hat{A}_1 = \frac{1}{\sqrt{|P_3|}}X_{P_1,P_3},$ $\hat{A}_2 = \frac{1}{\sqrt{|P_3|}}X_{P_2,P_3}$ \label{step:2_csw} \;
Compute vector $\hat{m}_1 = \frac{1}{|P_1|} \sum_{j \in P_1} X_{P_1,j}$ \;
Compute matrix $\hat{B} = \frac{1}{|P_4|} \sum_{j \in P_4} w_j X_{P_1,j} X_{P_2,j}^T$ \;
$\hat{\mu}_{P_1},\hat{\alpha}_1 \gets$ SearchSubroutine($\hat{A}_1,\hat{A}_2,\hat{B},\hat{m}_1,k$) \label{step:5_csw} \;
Compute $V_{P_1} = \{j \in P_1: \hat{\mu}_{P_1,j} > \tau \}$ \label{step:thresholding} \;
Repeat steps \ref{step:2_csw}-\ref{step:5_csw} with $P_i$'s rotated in order to estimate $\hat{\mu}_{P_2},\hat{\mu}_{P_3},\hat{\mu}_{P_4}.$ Use them to compute $V_{P_2},V_{P_3},V_{P_4}$ as in step \ref{step:thresholding} \;
Return community $\hat{V}_1 = V_{P_1} \cup V_{P_2} \cup V_{P_3} \cup V_{P_4}$\;
\caption{Community Search}
\label{alg:comm_search_white}
\end{algorithm}  
\DecMargin{1em}

\IncMargin{1em}
\begin{algorithm}[ht]
\SetAlgoLined
\KwIn{$\hat{A}_1,\hat{A}_2,\hat{B},\hat{m}_1,k$}
\KwOut{$\hat{\mu}_1, \hat{\alpha}_1$}
Compute rank $k$-svd of matrices $\hat{A}_1,\hat{A}_2:$ $\hat{A}_1 = U_1 D_1 V_1^T,$ $\hat{A}_2 = U_2 D_2 V_2^T$\;
Compute matrices $W_1 = U_1 D_1^{-1},$ $W_2 = U_2 D_2^{-1}$\;
Let $u_1$ be the largest left singular vector of $W_1^T \hat{B} W_2$\;
Compute $z = U_1 D_1 u_1$\;
Compute $a = u_1^T W_1^T \hat{m}_1$\;
Return $\hat{\mu}_1 \leftarrow z/a$ and $\hat{\alpha}_1 \leftarrow a^2$\;
\caption{SearchSubroutine}
\label{alg:whiten_subroutine}
\end{algorithm}  
\DecMargin{1em}

\section{Main Result} \label{sec:results}

In this section we present our main theoretical results. First we show that when the set of biased weights $(w_1,\hdots,w_n)$ satisfy certain mild sufficient conditions, then Algorithm \ref{alg:comm_search_white} is guaranteed to recover the target community $V_1.$ Later we show how such weights can be obtained even with a set of labeled nodes from the target community.

\subsection{Recovery using biased weights}

When side information is available in the form of biased weights $w_j$ for each node $j \in V,$ these weights need to be {\em informative} about the target community $V_1$ so that it could be recovered. Clearly {\em good} side information will lead to a better performance of any search algorithm. We quantify this quality of information in the following set of assumption (condition (A1) and (A2)) on the biased weights. The third condition (A3) is a more fundamental condition that determines when the community structure itself is identifiable in a stochastic block model.   

\begin{itemize}

\item {\em (A1) Average weight bias:} Under this condition the expected weight of a node in community $V_1$ is greater than the expected weight of a node in any other community $V_i.$ Precisely the weights satisfy:
$$
E[w_j|j \in V_1] > E[w_j | j \in V_i], \ \forall i \neq 1
$$
This weight bias allows us to determine that community $V_1$ is being searched / preferred over the remaining communities. However we only require this to hold in expectation and the actual weights themselves may vary significantly. Clearly any algorithm which only uses the weight bias to determine community membership by simple thresholding will perform very poorly. 

\item {\em (A2) Weight concentration:} Let $\alpha_{max} = \max_{i \in [k]} \alpha_i$ and $\alpha_{min}= \min_{i \in [k]} \alpha_i.$ Define $\sigma_1(R):=E[w_j | j \in V_1],$ $\sigma_2(R):= \max_{i \neq 1} E[w_j | j \in V_i],$ $\gamma_2:=\max_{i \in [k], j\in V} |w_j-E[w_j|j \in V_i]|,$ and $\xi(n)=o(\sqrt{\log n})$ be any slowly growing function. Then with high probability the maximum deviation of the weights are bounded as,
\begin{equation*}
 \frac{\gamma_2}{(\sigma_1(R)-\sigma_2(R))} = O\left( \min \left\{ \frac{\alpha_{min}^4(p-q)^4}{\alpha_{max}^4 p^4 \xi(n)}, \frac{\alpha_{min}^5 \sqrt{n} (p-q)^5}{\alpha_{max}^4 p^{4.5} \xi(n)}-1\right\}  \right)
\end{equation*}
This condition dictates that the maximum variation of the weights $\gamma_2$ is also small compared to the difference between the largest and second largest expected weights $\sigma_1(R)-\sigma_2(R).$ Since the weights are used primarily to construct the matrix $B$ in Algorithm \ref{alg:comm_search_white}, this condition ensures that the matrix $B$ can be estimated up to a tolerable error.

\item {\em (A3) $p,q$ separation:} Let $p,q,n$ satisfy
$$
\frac{(p-q)^2}{p\sqrt{p}} = \tilde{\Omega}\left( \frac{\alpha_{max}}{\alpha_{min}^2 \sqrt{n}} \right)
$$
This condition fundamentally determines when communities are identifiable in a stochastic block model and similar conditions are required for other community detection algorithms \cite{AndGeHsuKak13,ChaChuTsi12,ChenSanXu12}. The more the gap $p-q$ easier it is to identify communities. Hence this condition gives a lower bound on $p-q$ which is required for community identifiability.
\end{itemize}

Theorem \ref{thm:whiten_oracle} shows that under the above assumptions on the biased weights Algorithm \ref{alg:comm_search_white} can reconstruct community $V_1$ with high accuracy. 

\begin{theorem} \label{thm:whiten_oracle}
Consider a $(n,k,p,q)$ stochastic block model satisfying condition (A3). Given biased weights $(w_1,\hdots , w_n)$ satisfying conditions (A1), (A2), then Algorithm \ref{alg:comm_search_white} recovers community $V_1$ with fraction of error nodes $o(1)$ with high probability.
\end{theorem}

\begin{remark}
For a stochastic block model with equal community sizes $n/k$ condition (A3) reduces to $\frac{(p-q)^2}{p\sqrt{p}}=\tilde{\Omega}\left(\frac{k}{\sqrt{n}} \right).$ When $p=\Theta(p-q)$ this has the same scaling as other community detection algorithms \cite{ChenSanXu12,AndGeHsuKak13,ChaChuTsi12}. Therefore even in sparse graphs where $p,q = \Theta\left(\frac{\log n}{n} \right)$ or for small community sizes up to $\Omega(\sqrt{n})$ nodes Algorithm \ref{alg:comm_search_white} can recover the community. 
\end{remark}

In Theorem \ref{thm:whiten_oracle} the $o(1)$ fraction error can be easily converted to a zero error guarantee using an additional post-processing step. Instead of estimating community $1$ nodes inside partition $P_1$ we can estimate those in partition $P_2,$ first by observing for each node $j\in P_2$ the number of edges shared with the estimated set $V_{P_1},$ followed by thresholding. Since $V_{P_1}$ estimates $V_1 \cap P_1$ up to only $o(\alpha_1 n)$ error nodes this does not cause any errors in thresholding, with high probability. This post-processing step is also independent of the previous steps in the algorithm since the edges between partitions $P_1$ and $P_2$ are not utilized in Algorithm \ref{alg:comm_search_white}. The following theorem formalizes this idea.

\begin{theorem}[Exact recovery] \label{thm:search_exact_recovery}
In a $(n,k,p,q)$ stochastic block model, under assumptions (A1)-(A3), Algorithm \ref{alg:comm_search_white} with an additional degree thresholding step can recover community $V_1$ completely with high probability.
\end{theorem} 

We prove Theorems \ref{thm:whiten_oracle} and \ref{thm:search_exact_recovery} in Appendix \ref{app:whiten_proof}. 

\subsection{Recovery using labeled nodes} \label{sec:recovery_labeled_node}

Biased weights, as required in Theorem \ref{thm:whiten_oracle}, can be obtained from a small set of labeled nodes $\mathcal{L}$ as follows:

\begin{itemize}
\item Choose a radius $r$
\item Weight $w_i$ is the number of edges between nodes in $\mathcal{L}$ and nodes at a distance of $r$ hops from node $i$
\end{itemize}

Note that the weight can also be viewed as the number of neighbors of the set $\mathcal{L}$ which are at a distance $r$ from node $i.$ Larger choice of radius $r$ means less variance in the weights, but also potentially less bias if it becomes too large. For example, $r=1$ means only neighbors of labeled nodes get weights; this is very high bias but also high variance.


The theorem below provides the correct way to choose the radius $r$ such that the weights $w_i$ can be made to satisfy conditions (A1), (A2). This means that even with such weights computed via labeled nodes we can efficiently find community $V_1$ using Algorithm \ref{alg:comm_search_white}. Note that when $p\geq \frac{1}{\sqrt{n}}$ then with high probability the labeled nodes in $\mathcal{L}$ has neighbors with any other node $i \in V_1 \backslash \mathcal{L},$ hence the number of common neighbors between $i$ and nodes $l \in \mathcal{L}$ can be taken as weights $w_i$ which will satisfy conditions (A1), (A2). However this does not work for sparse graphs when $p<\frac{1}{\sqrt{n}}.$ In the following theorem we show that even for $p=\Theta\left(\frac{\log n}{n^\epsilon} \right),$ $\frac{1}{2}\leq\epsilon\leq 1$ the weights chosen by the above procedure and a correct $r$ will work.

\begin{theorem} \label{thm:labeled_nodes}
Consider a $(n,k,p,q)$ stochastic block model satisfying condition (A3) where $p=\Theta\left(\frac{\log n}{n^\epsilon}\right),$ $q=\Theta\left(\frac{\log n}{n^\epsilon}\right),$ $p-q=\Theta\left(\frac{\log n}{n^\epsilon}\right)$ and all equal sized communities, where $\frac{1}{2}\leq\epsilon\leq 1.$ Given $L=\tilde{\Omega}(n^{\epsilon/2})$ labeled nodes, the biased weights computed with $r = \frac{2\log (n^\epsilon /L) }{\log np},$ satisfy conditions (A1), (A2) with high probability.
\end{theorem}

We prove this in Appendix \ref{app:labeled_node_proof}. For simplicity in Theorem \ref{thm:labeled_nodes} we assume equal community sizes, however this can be extended to unequal but comparable community sizes.

\subsection{Parallel semi-supervised graph clustering} \label{sec:parallel_clustering}

Our algorithm naturally provides a method for the standard semi-supervised graph clustering problem. This is the setting where we are given a small number of labeled nodes from every community, and we are interested in recovering all communities. In such a scenario we can apply the community search algorithm to search for each individual community using the labeled nodes in that target community. Moreover this search can be performed in parallel. Therefore Algorithm \ref{alg:comm_search_white} can also be used as a parallel graph clustering algorithm. Note that the vector $m_1$ and matrices $A_1,A_2$ remain the same for individual searches, only matrix $B$ should be computed separately for every target community. Section \ref{sec:experiments} shows some numerical results evaluating the performance of Algorithm \ref{alg:comm_search_white} in this semi-supervised graph clustering setting.

\subsection{Comparison} \label{sec:comparison}

In this section we compare the theoretical performance of our algorithm with other unsupervised graph clustering algorithms.

For graphs with equal communities of size $n/k,$ convex optimization based algorithms by \cite{ChenSanXu12,AilChenXu13,AgarBanKoiKol:15sdp} can achieve the performance bound $\frac{(p-q)}{\sqrt{p}}=\tilde{\Omega}\left(\frac{k}{\sqrt{n}}\right).$ In comparison our algorithm achieves a slightly higher bound $\frac{(p-q)^2}{p\sqrt{p}}=\tilde{\Omega}\left(\frac{k}{\sqrt{n}}\right).$ However when $p=\Theta(p-q)$\footnote{This is the case in most real sparse networks when $p=\Theta(\log n/n),$ if not then it becomes impossible for any algorithm to recover communities in this regime as shown by \cite{AbbeBanHall:14}.} both bounds are equivalent (up to log factors) implying our algorithm can recover communities even in sparse graphs with $p,q=\Theta\left(\frac{\log n}{n}\right)$ and for growing number of communities $k=O(\sqrt{n}).$ In terms of runtime our algorithm runs in $O(n^2k)$ time faster than $\Omega(n^3)$ time required by convex optimization based algorithms.

The Community Search by Whitening algorithm is also faster than tensor decomposition based graph clustering algorithm by \cite{AndGeHsuKak13}. Note that the first step of this tensor algorithm is to compute a whitening matrix using rank--$k$ svd, which is identical to the search algorithm. In the remaining steps, for the tensor algorithm, the bulk of the computation is a rank--$k$ tensor decomposition requiring $O(k^5)$ computation, which is slower than rank--$1$ svd computed in $O(k^2)$ time by the search algorithm. This is corroborated by our experiments in Section \ref{sec:experiments}.

Recently a quasi--linear time graph clustering algorithm was presented by \cite{AbbeSan:15} for the case when number of communities $k=O(1)$. In comparison our algorithm can be applied even when the number of communities scale as $k=O(\sqrt{n}),$ and it requires much lesser knowledge of model parameters than the former.

\section{Experiments} \label{sec:experiments}

In this section we present our numerical results showing the
performance of the Community Search algorithm on synthetic and real
datasets. We compare our algorithm with the Spectral clustering
algorithm by Ng et al. \cite{NgJorWei02} and the Tensor decomposition
based clustering algorithm by Anandkumar et al. \cite{AndGeHsuKak13}. We generate synthetic datasets according to
the stochastic block model (see Section \ref{sec:algo}) with $n=1000$
nodes, $k \in \{5,8\}$ communities, and different values of $p$ and
$q.$ The real world network we consider is the US political blogosphere network first introduced in \cite{AdaGla:05blogosphere}. The Spectral clustering algorithm \cite{NgJorWei02} requires
clustering of the rows corresponding to bottom $k$ eigenvectors of the
normalized Laplacian. Although k-means may be used for this, it tends
to converge to local minima resulting in poor performance. To prevent
this we perform clustering of the rows via the hierarchical SLINK
algorithm \cite{slinkSib73}. We refer to this Spectral+SLINK algorithm
simply as Spectral clustering in the remaining section. Our algorithm
implementations are all in Matlab.
%
We consider two types of side information:
{\em labeled nodes} and {\em synthetic weights.} 

\noindent \textbf{Labeled Nodes:} As discussed earlier, this is a natural means
of providing side information to the algorithm. A set of $m$ labeled
nodes are randomly chosen from the target community $V_1.$ The
corresponding weights are then computed as described in Section
\ref{sec:recovery_labeled_node} with $r \in \{1,2\}.$

\noindent \textbf{Synthetic Weights:} We synthetically generate three sets of
weights, each of which are (on average) larger over the target
community. These weights are generated as follows. A pair of weights
$(w_1, w_2)$ are first chosen to be one of $\{(5,8), (5, 10), (5, 12)\}.$
For each node in community $V_1,$ we set the node's weight to be $w_2$
with probability $0.8,$ and $w_1$ other-wise. For all other nodes in
$V\backslash V_1,$ we swap the probabilities, i.e., we set a node's weight to
be $w_2$ with probability $0.2$ and $w_1$ other-wise.  This process
generates three possible values of the expected node weights in the
target community, $\sigma_1(R) \in \{7.4,9,10.6\}.$




\noindent\textbf{Performance Metrics:} Note that our algorithm directly uses labeled nodes/biased node weights, and the graph to infer the target community. The baseline algorithms however first estimate all communities in the graph, then it computes the average node weight in each community, finally outputs as target the community which has the highest average node weight. The estimation error for the $i$-th
community is given as
$e_i=|\{j \in V: j \in V_i, j \not \in \hat{V}_i \ or \ j \in
\hat{V}_i, j \not \in V_i\}|.$
We compute the error for searching each community and plot either the
overall average, or average over a subset of clusters. Let $T_{\mathcal{A}_1},T_{\mathcal{A}_2}$ be the runtimes of algorithms $\mathcal{A}_1,\mathcal{A}_2$ respectively. Then we define speedup $s$ of algorithm $\mathcal{A}_1$ over algorithm $\mathcal{A}_2$ as $s=T_{\mathcal{A}_2}/T_{\mathcal{A}_1}.$ $s>1$ implies algorithm $\mathcal{A}_1$ is faster than $\mathcal{A}_2.$   

\begin{figure}[hptb]
\centering
\subfigure[]{\includegraphics[height=1.8in]{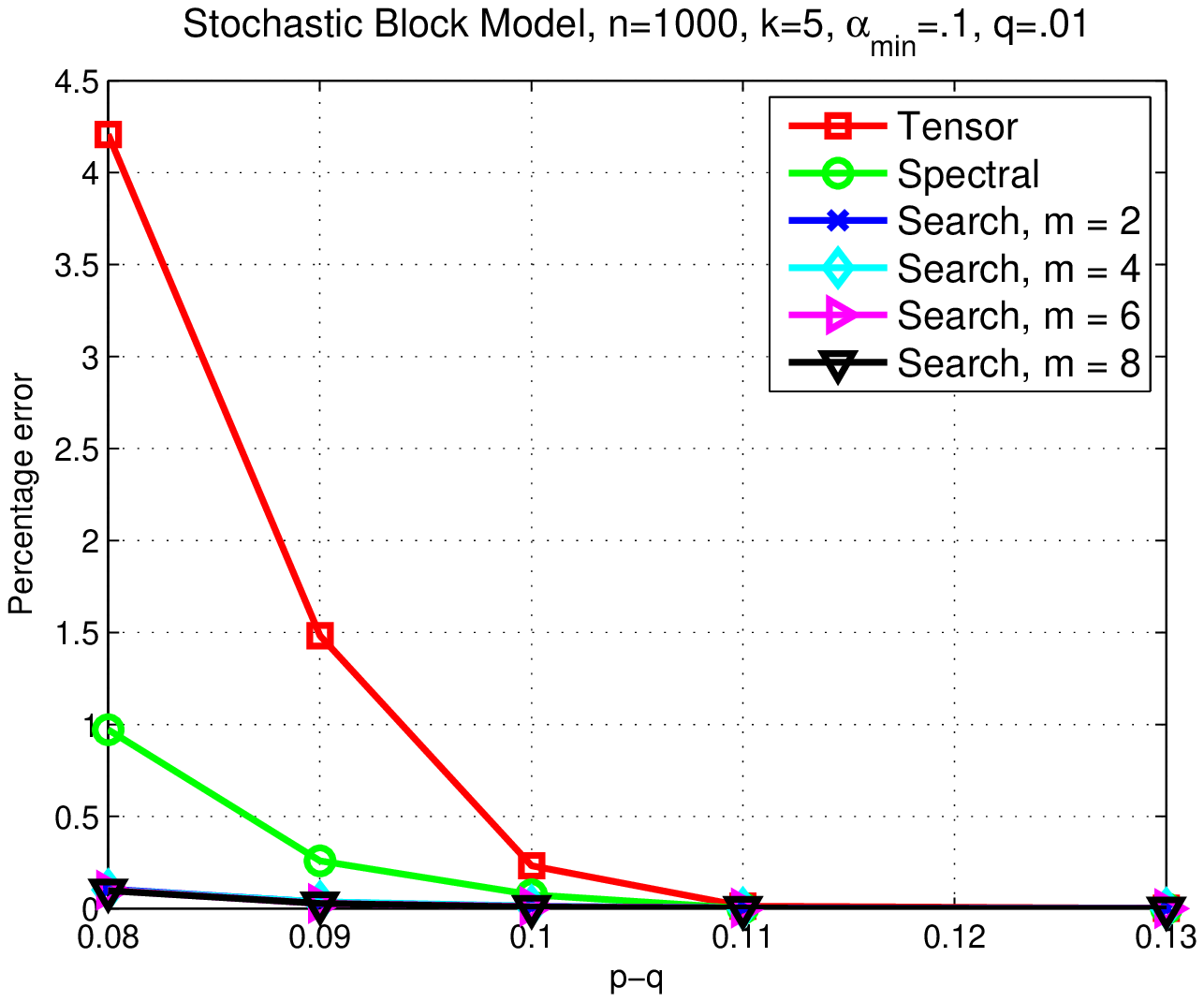}}
\subfigure[]{\includegraphics[height=1.8in]{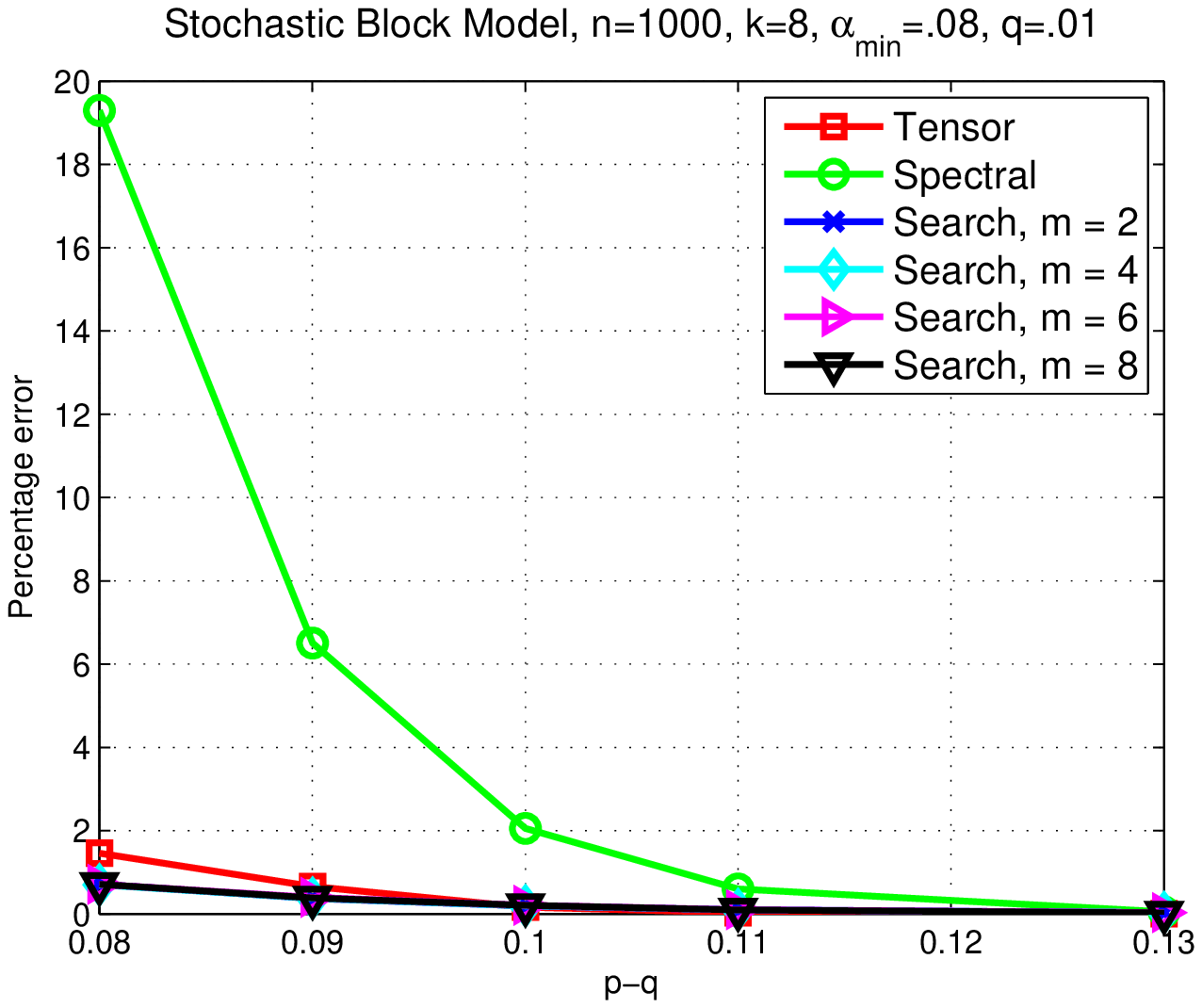}}
\caption{Labeled Nodes: Comparing the average error performance
  of Community Search algorithm with Spectral clustering
  \cite{NgJorWei02} and Tensor decomposition \cite{AndGeHsuKak13}
  algorithms in a stochastic block model with (a) $n=1000,$ $k=5,$
  $\alpha_{min}=.1,$ (b) $n=1000,$ $k=8,$
  $\alpha_{min}=.08.$ The algorithms use $m$ labeled node from target
  cluster as side information and compute biased weights. The
  Community Search algorithm outperforms both Spectral clustering and
  Tensor decomposition.}
\label{fig:complexityLabeledWt}
\end{figure}

\subsection{Performance and Speedup with Labeled Nodes}
\label{ssec:sim-nodes}

First we compare the error performance of Community Search algorithm
with Spectral clustering and Tensor decomposition algorithms in the
setting where side information is given in the form of $m$ labeled
nodes from the target community. We then compute biased weights $w_j$
using the tree method of Section \ref{sec:recovery_labeled_node} with
a radius $r=2.$ Note that this tree method may assign weights in
violation of condition (A1) for small target communities, since for
small target clusters the number of nodes in the tree from a large
cluster may exceed those from the target community, in such cases Algorithm \ref{alg:comm_search_white} cannot be expected to recover the communities. Therefore we consider a subset of larger communities which assign the
correct weights satisfying condition (A1) and evaluate our algorithm over these communities. Figure
\ref{fig:complexityLabeledWt} (a) plots the average error over $3$
largest cluster in a stochastic block model (SBM) with $n=1000,$ and
$k=5$ unequal sized communities. The Community Search shows
significantly less error than Tensor decomposition and Spectral
clustering. In Figure \ref{fig:complexityLabeledWt} (b) we plot the
average over $5$ larger cluster in a SBM with $n=1000,$ $k=8$ unequal
communities. Again Community Search shows better error than Spectral
clustering and comparable error to Tensor decomposition.

\begin{figure}[hptb]
\centering
\subfigure[]{\includegraphics[height=1.8in]{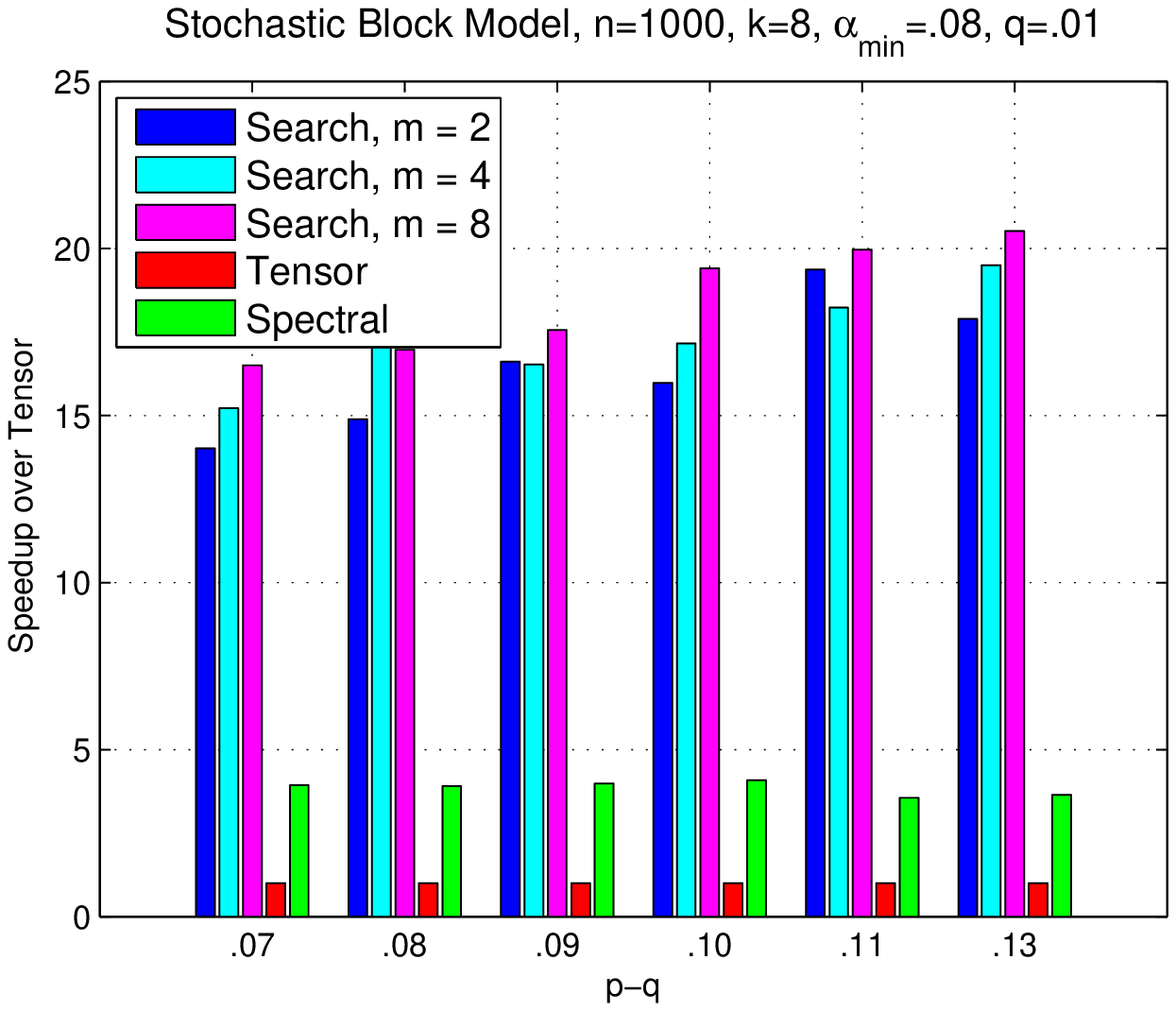}}
\subfigure[]{\includegraphics[height=1.8in]{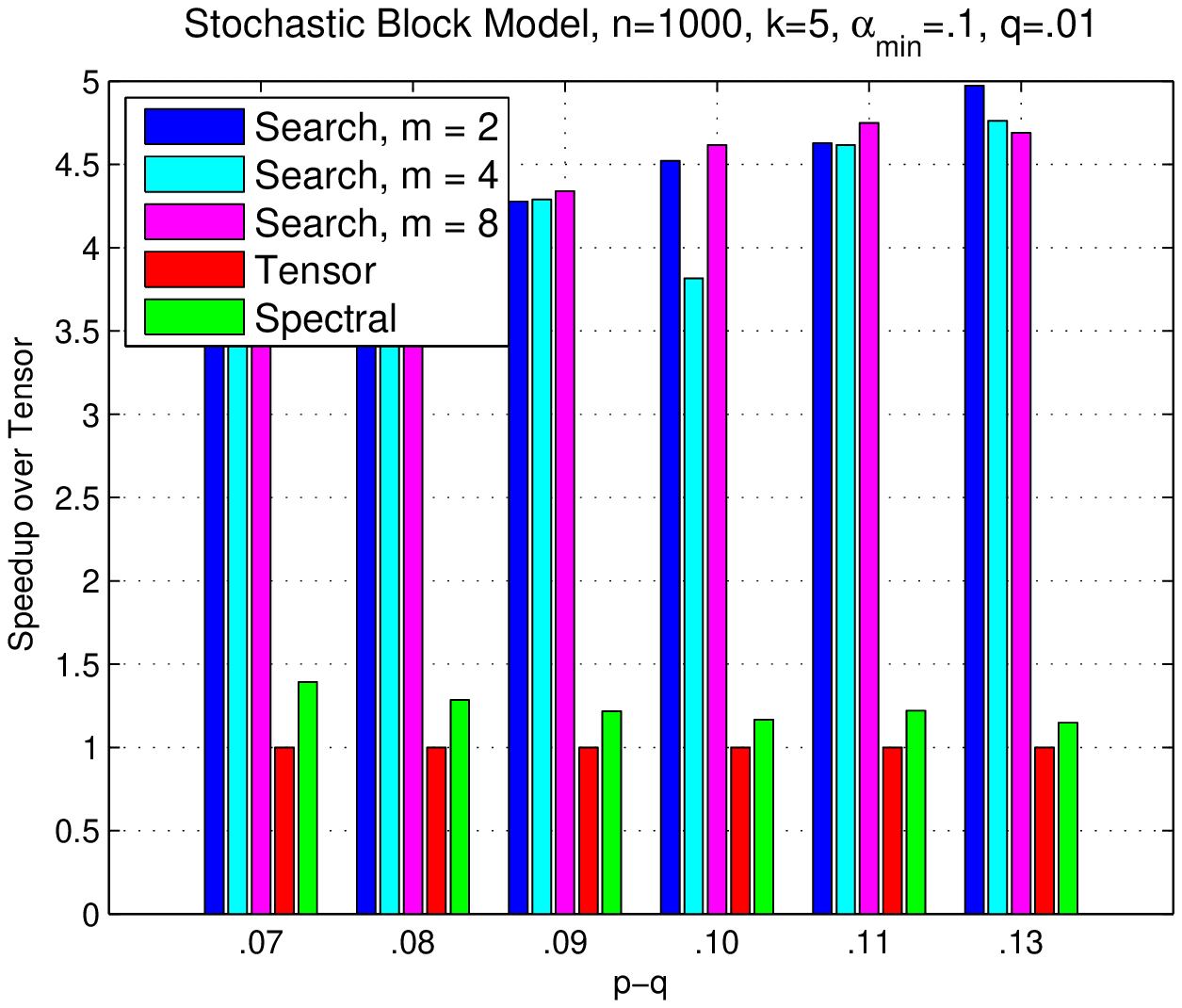}}
\caption{Labeled Nodes: The average speedup performance
  of the Community Search and Spectral clustering \cite{NgJorWei02}
  algorithms with respect to Tensor decomposition \cite{AndGeHsuKak13}
  in a stochastic block model with (a) $n=1000,$ $k=8,$
  $\alpha_{min}=.08,$ (b) $n=1000,$ $k=5,$ $\alpha_{min}=.1,$ and labeled nodes as side information. The
  Community Search algorithm is faster than both Spectral clustering
  and Tensor decomposition.}
\label{fig:speedupLabeledWt}
\end{figure}

Figure \ref{fig:speedupLabeledWt} show the speedup performance of the Community Search and Spectral
clustering algorithms over Tensor decomposition in this setting with
labeled nodes. As indicated earlier, all three algorithms were
implemented in Matlab. We observe that the Community Search has a much
lower runtime than both Spectral clustering and Tensor decomposition.

\subsection{Performance and Speedup with Synthetic Weights}
\label{ssec:sim-bw}

\begin{figure}[hptb]
\centering
\subfigure[]{\includegraphics[height=1.8in]{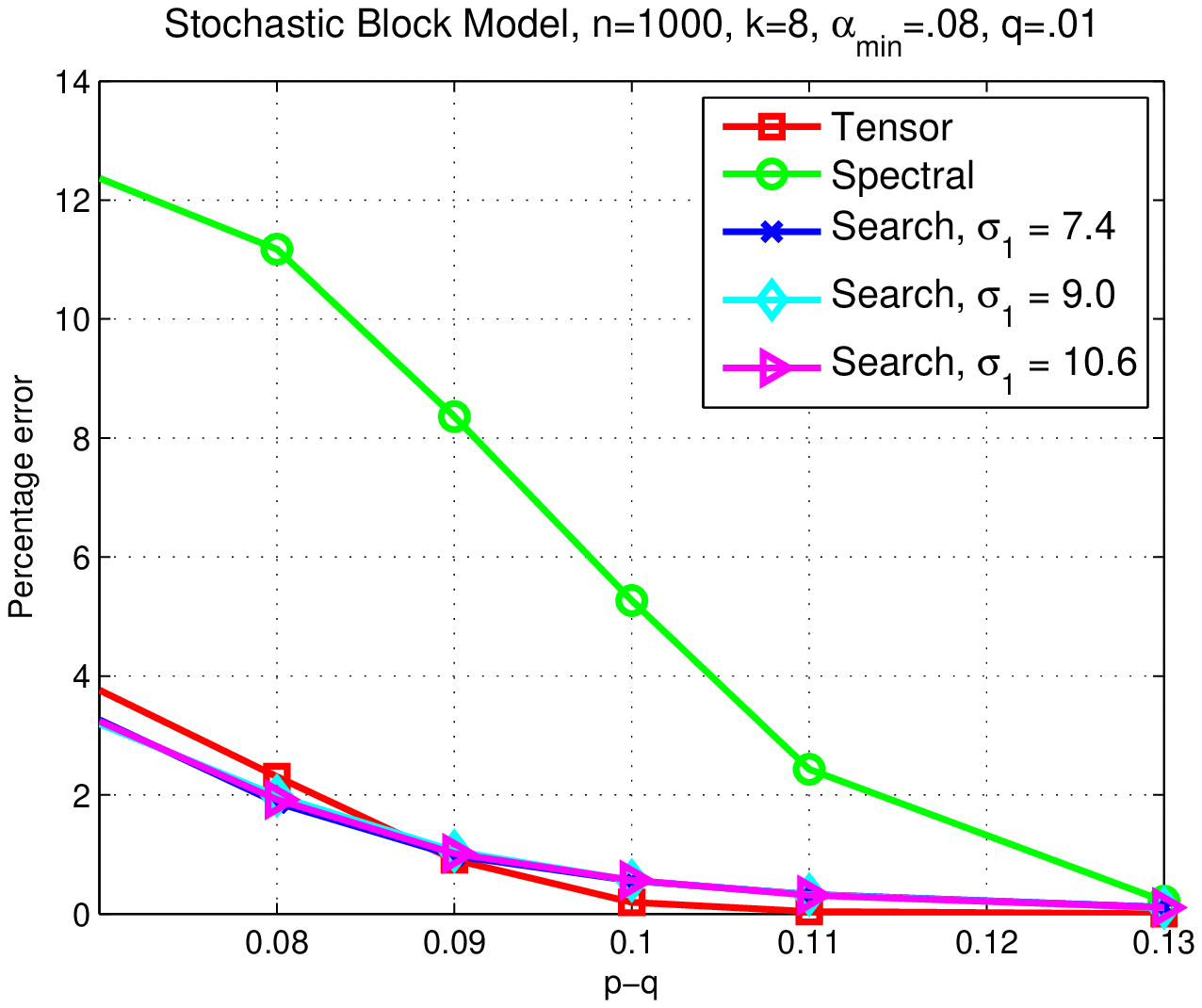}}
\subfigure[]{\includegraphics[height=1.8in]{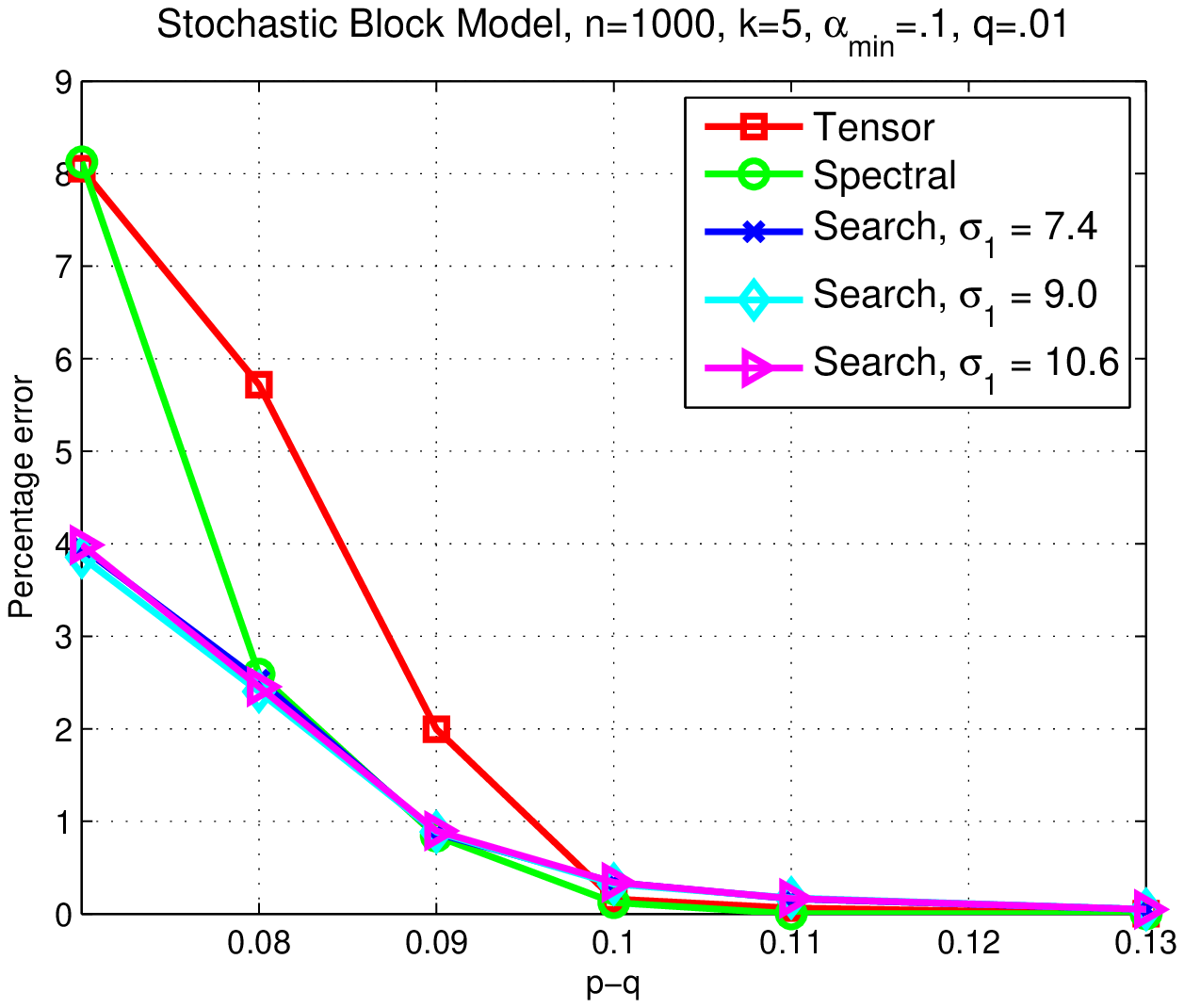}}
\caption{Synthetic Weights: Comparing the average error performance of Community
  Search algorithm with Spectral clustering \cite{NgJorWei02} and
  Tensor decomposition \cite{AndGeHsuKak13} algorithms in a stochastic
  block model with (a) $n=1000,$ $k=8,$ $\alpha_{min}=.08,$ (b) $n=1000,$ $k=5,$
  $\alpha_{min}=.1.$ The algorithms
  use synthetic weights as side information to search for the target
  community. Community Search algorithm shows a lower error.}
\label{fig:complexityOracleWt}
\end{figure}

Next we compare the error and runtime performance of all three
algorithms in a setting where side information is available in the
form of synthetically generated biased weights (as discussed earlier,
three different choices of parameters). Figure
\ref{fig:complexityOracleWt} (a) plots the average error over all
communities in a SBM with $n=1000,$ $k=8.$ The Community Search
algorithm has a better performance over Spectral clustering and
comparable performance with Tensor decomposition. In Figure
\ref{fig:complexityOracleWt} (b) plots the average error in a SBM with
$n=1000,$ $k=5.$ In this case Community Search outperforms Tensor
decomposition and has comparable performance to Spectral clustering.

\begin{figure}[hptb]
\centering
\subfigure[]{\includegraphics[height=1.8in]{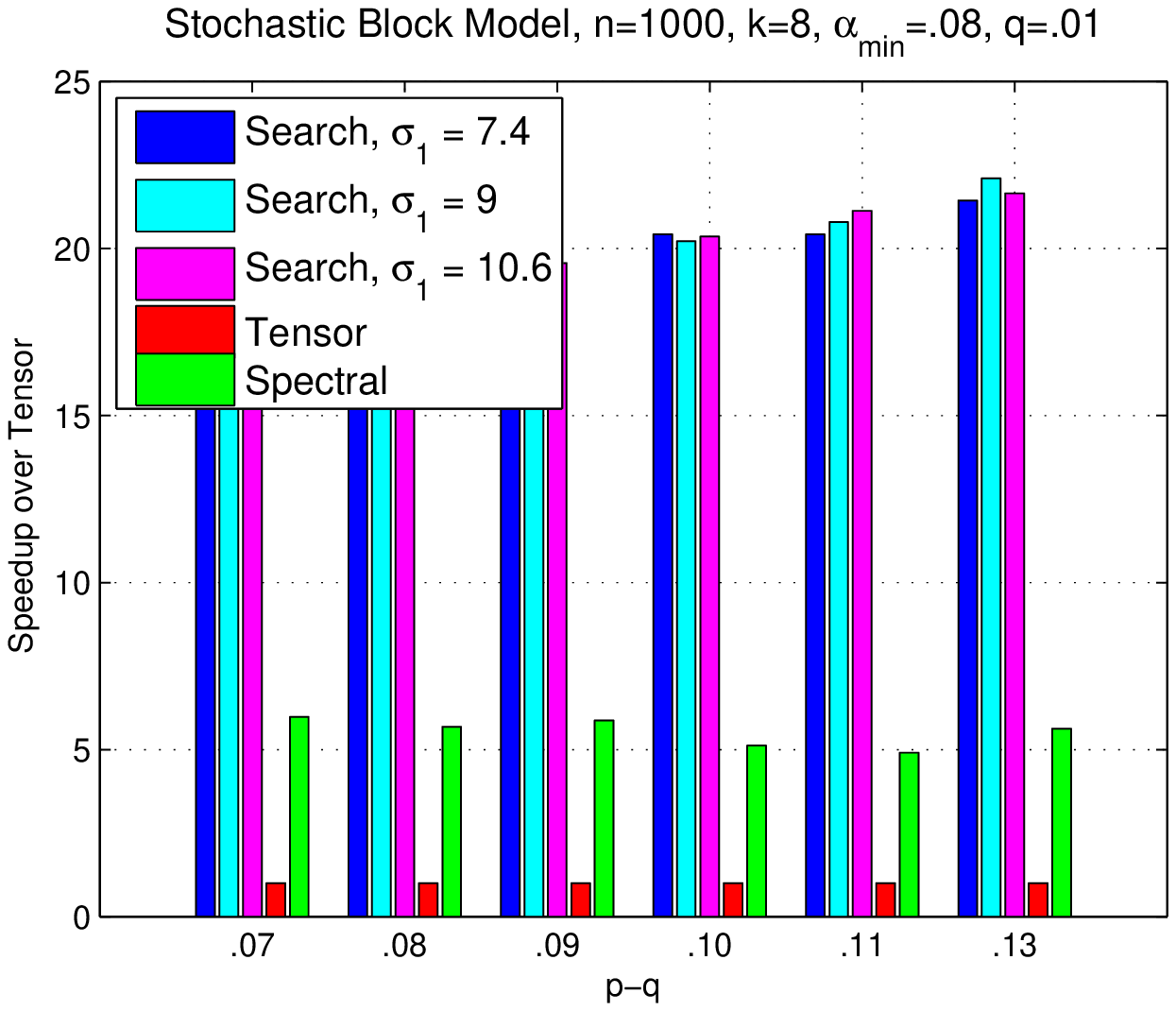}}
\subfigure[]{\includegraphics[height=1.8in]{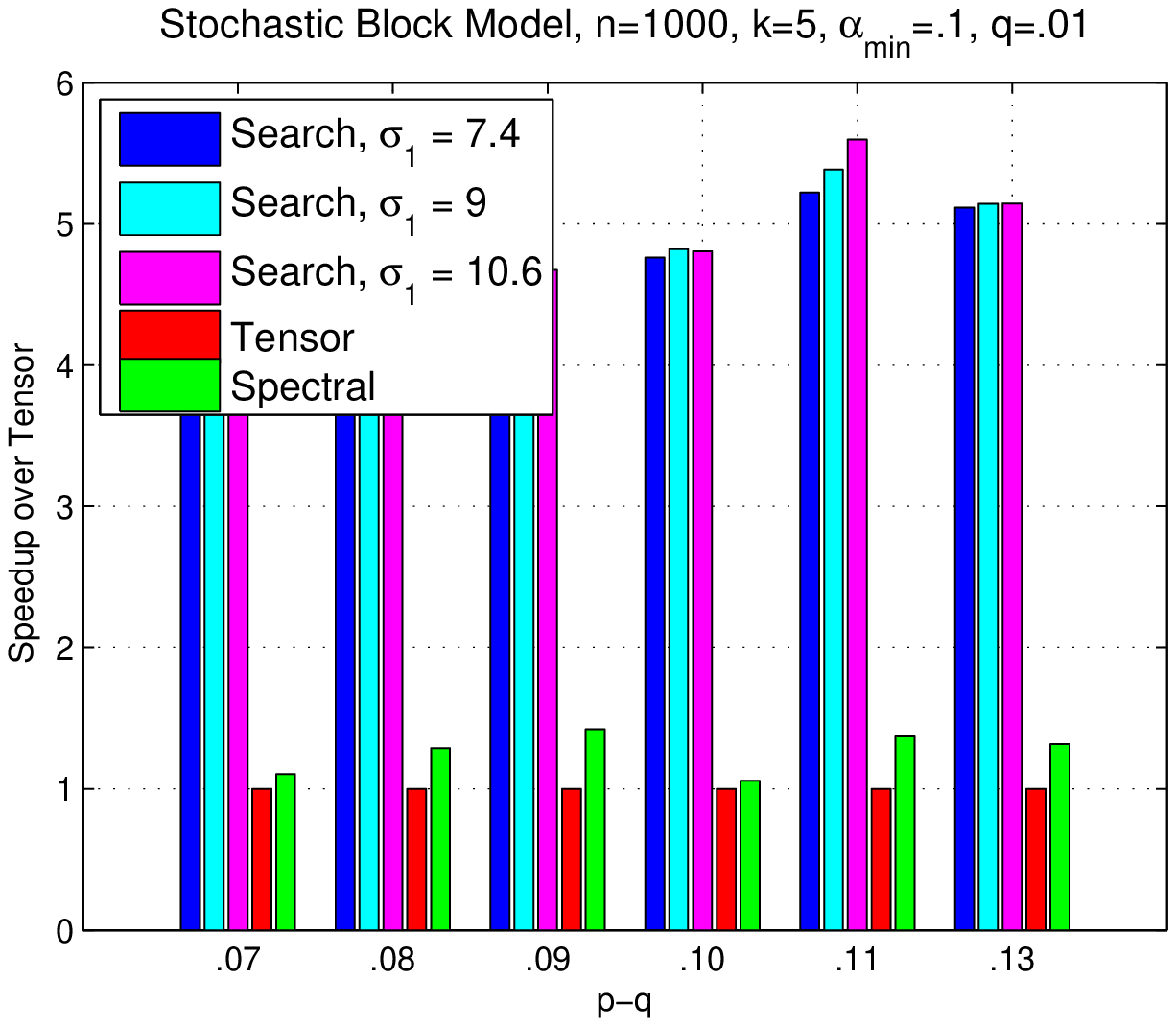}}
\caption{Synthetic Weights: The average speedup
  performance of the Community Search and Spectral clustering
  \cite{NgJorWei02} algorithms with respect to Tensor decomposition
  \cite{AndGeHsuKak13} in a stochastic block model with (a) $n=1000,$
  $k=8,$ $\alpha_{min}=.08,$ (b) $n=1000,$
  $k=5,$ $\alpha_{min}=.1,$ and synthetic weights as
  side information. The Community Search algorithm is faster than both
  Spectral clustering and Tensor
  decomposition.}
\label{fig:speedupOracleWt}
\end{figure}

In Figure \ref{fig:speedupOracleWt} we plot the average speedup of Community Search and Spectral
clustering over Tensor decomposition. Again the Community Search
algorithm is significantly faster than both Spectral clustering and
Tensor decomposition. We also observe that the speedup increases with
increasing $p-q.$

\subsection{Sensitivity}
\label{ssec:sim-sensitivity}

In order to determine the sensitivity of our algorithm with respect to the quality of side information, and the number of communities $k,$ we perform the following two experiments.

\begin{figure}[hptb]
\centering
\subfigure[]{\includegraphics[height=1.8in]{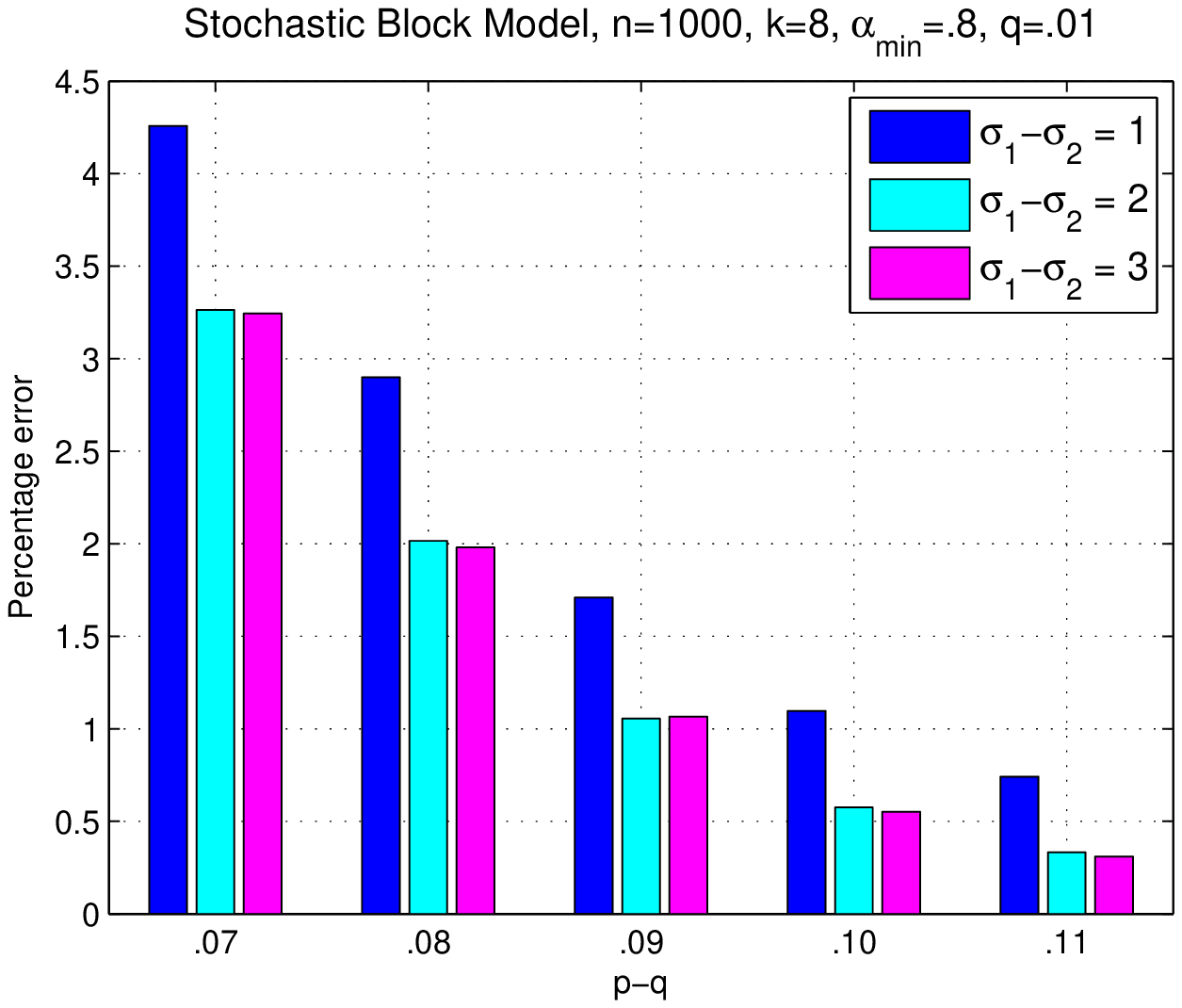}}
\subfigure[]{\includegraphics[height=1.8in]{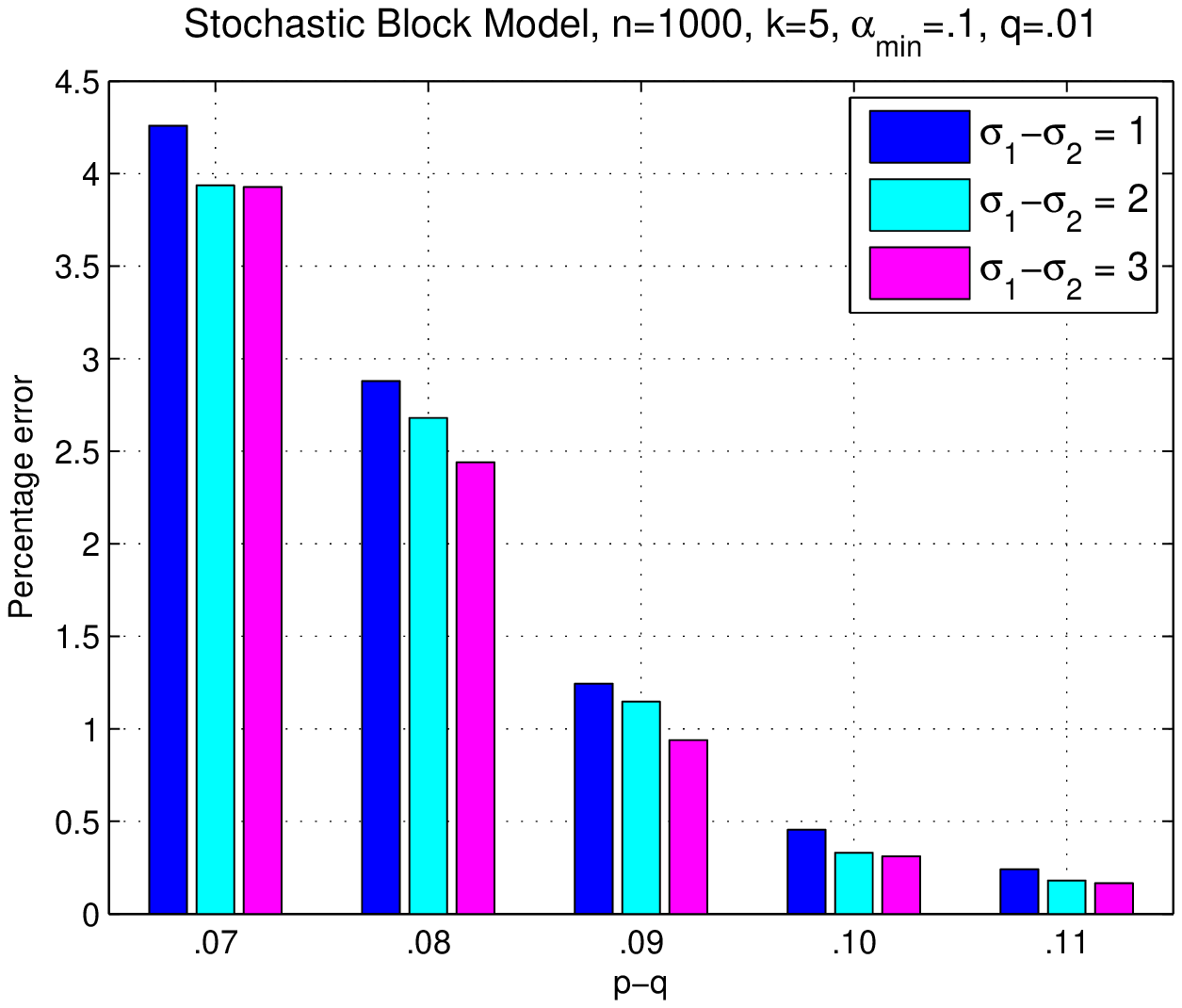}}
\caption{Sensitivity to side information: The average percentage error of Community Search Algorithm in a stochastic
  block model with (a) $n=1000,$ $k=8,$ $\alpha_{min}=.08,$ (b) $n=1000,$ $k=5,$
  $\alpha_{min}=.1,$ with increasing singular value gap $\sigma_1(R)-\sigma_2(R).$ As shown in our analysis the performance improves with increase in the singular value gap.}
\label{fig:sensitivityOracleWt}
\end{figure}

First, to see how the quality of side information effects the performance of our algorithm we plot the average error with increasing singular value gap $\sigma_1(R)-\sigma_2(R)$ (or the difference between the largest and second largest expected node weight) in Figure \ref{fig:sensitivityOracleWt}. In this experiment we fix the synthetic weights $(w_1=5,w_2=10)$ and vary $\sigma_1(R)-\sigma_2(R)$ by changing the probabilities with which the weights appear in each community. Note that the singular value gap increases when one weight appears with greater chance than the other. Therefore $\sigma_1(R)-\sigma_2(R)$ can also be viewed as a measure of quality of side information. As predicted from our analysis, we observe that the error improves with an increase in the gap $\sigma_1(R)-\sigma_2(R).$

\begin{figure}[hptb]
\centering
\subfigure[]{\includegraphics[height=1.8in]{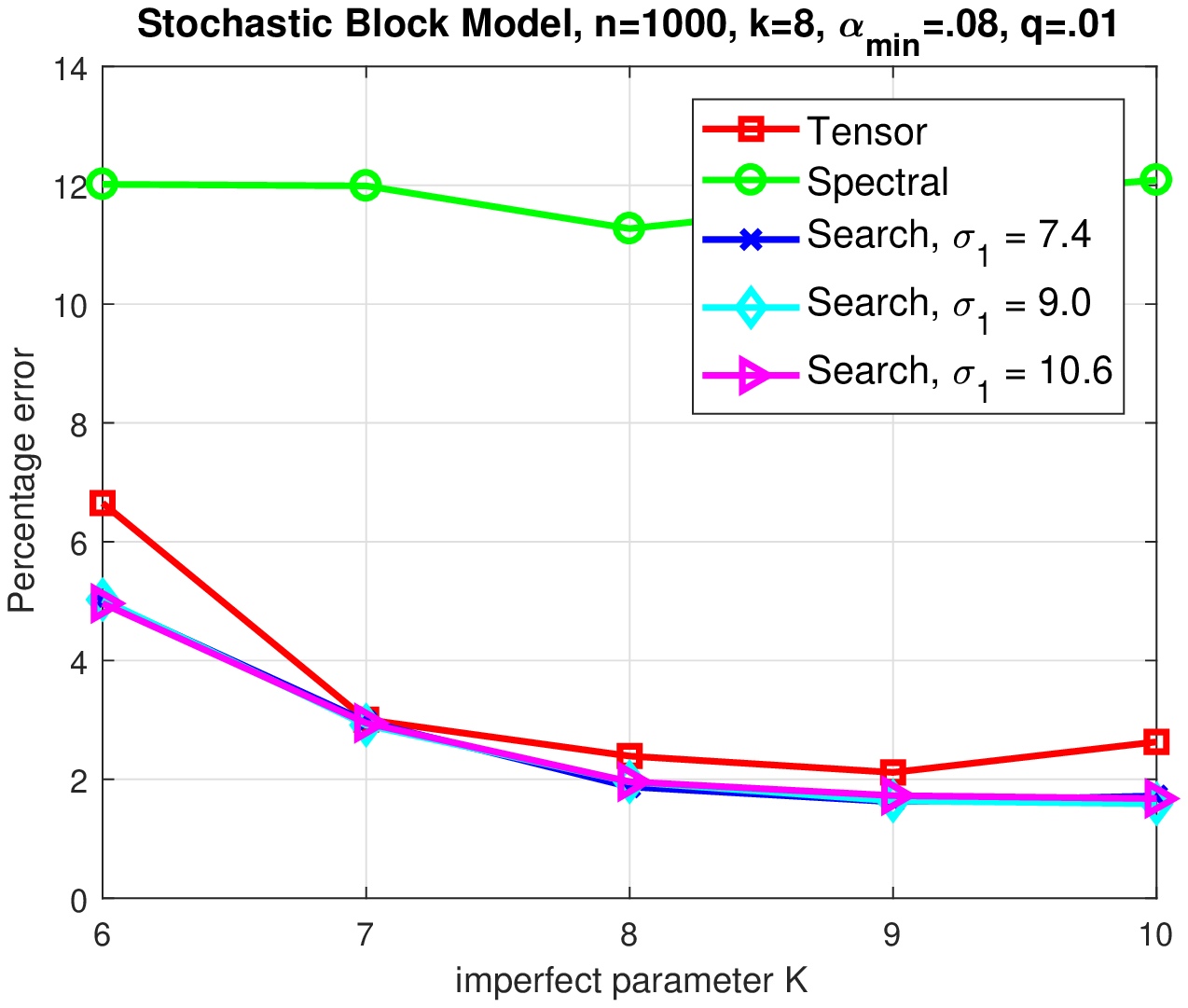}}
\subfigure[]{\includegraphics[height=1.8in]{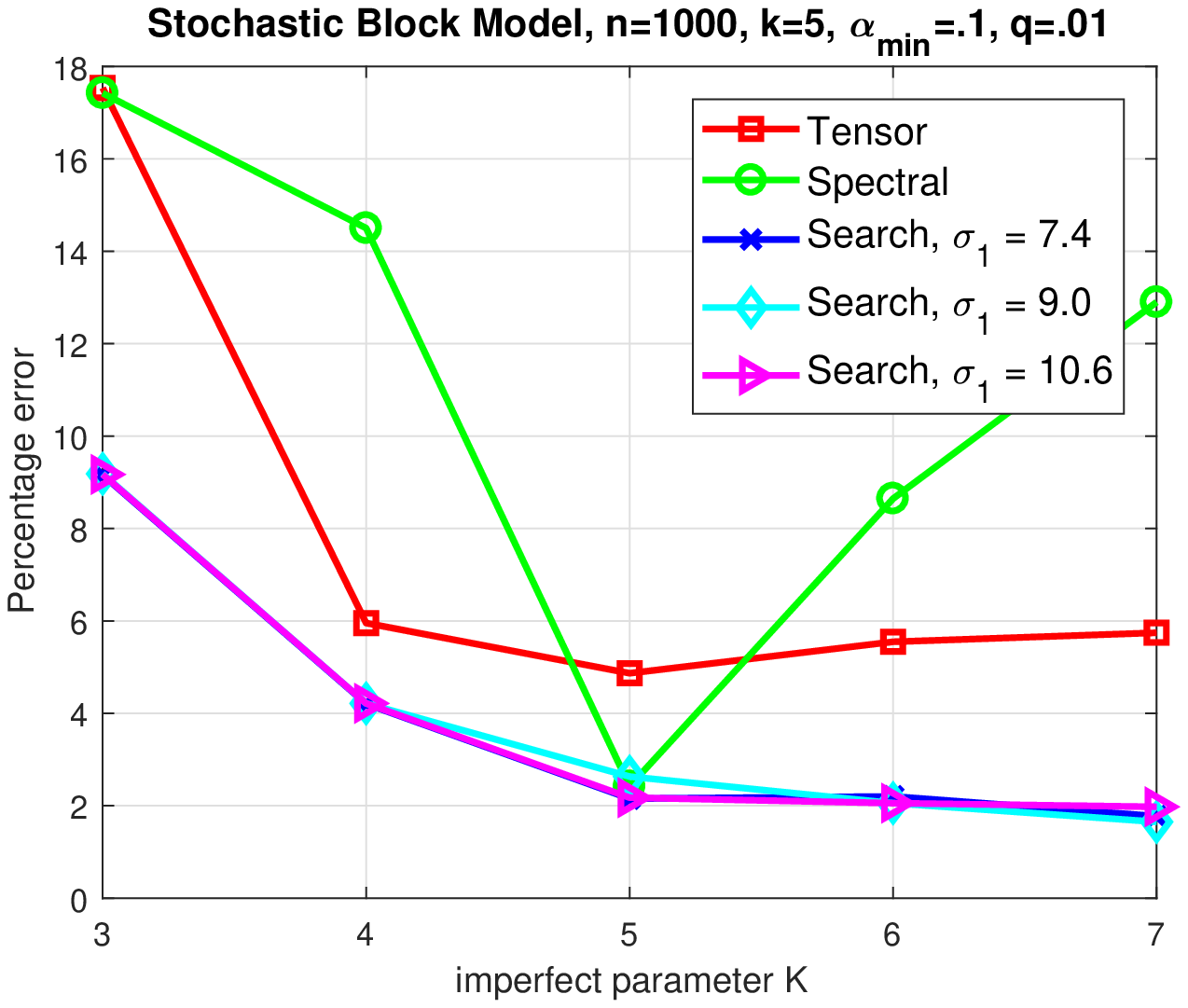}}
\caption{Sensitivity to $k$: The average percentage error of Community Search algorithm in a stochastic
  block model with (a) $n=1000,$ $k=8,$ $\alpha_{min}=.08,$ (b) $n=1000,$ $k=5,$
  $\alpha_{min}=.1,$ $p=.09, q=.01$ in both models, synthetic weights as side information, and with imperfect knowledge of number of communities $k.$ Our Community Search algorithm exhibits lower sensitivity and error than Spectral clustering
  \cite{NgJorWei02} and Tensor decomposition
  \cite{AndGeHsuKak13} algorithms for both models.}
\label{fig:sensitivityKOracleWt}
\end{figure}

Often in real applications one do not have perfect knowledge of the number of communities $k$ in a graph, which is required to run any community search or detection algorithm. Thankfully, there are several methods to estimate the parameter $k$ e.g. from the spectral properties of the graph \cite{ChenSanXu12,NewRei:16}. Another approach is to compute a suitable community quality score metric like modularity \cite{New06mod,YangLes12} after running the algorithm with different values of $k,$ and choosing the $k$ which produce a community with the best score. However, such estimation may not be always accurate. Therefore it is crucial that any community search algorithm perform robustly with respect to the input parameter $k$ in the algorithm. In our next experiment we compare the sensitivity of our Community Search algorithm with Spectral clustering and Tensor decomposition when provided with imperfect parameter $k.$ In Figure \ref{fig:sensitivityKOracleWt} we plot the average percentage error of all three algorithms on two different SBM. We observe that even with imperfect knowledge of $k$ the Community Search algorithm has lower error than Spectral clustering and Tensor decomposition. Interestingly, the Community Search algorithm shows much less sensitivity to higher $k$ values than a lower $k$ (with respect to the ground truth $k$).

\subsection{Parallel Clustering}
\label{ssec:sim-parallel}

\begin{figure}[hptb]
\centering
\subfigure[]{\includegraphics[height=1.8in]{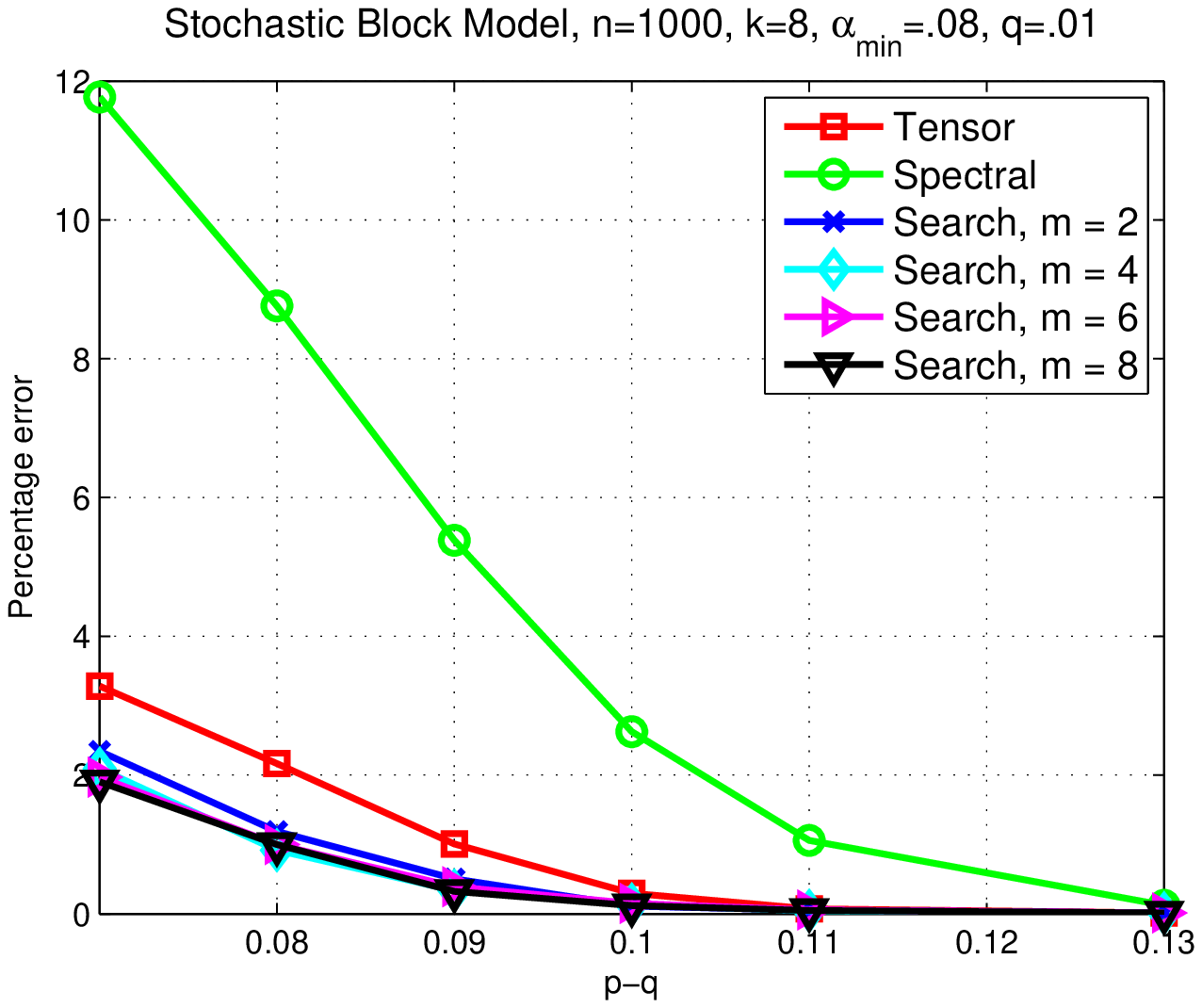}}
\subfigure[]{\includegraphics[height=1.8in]{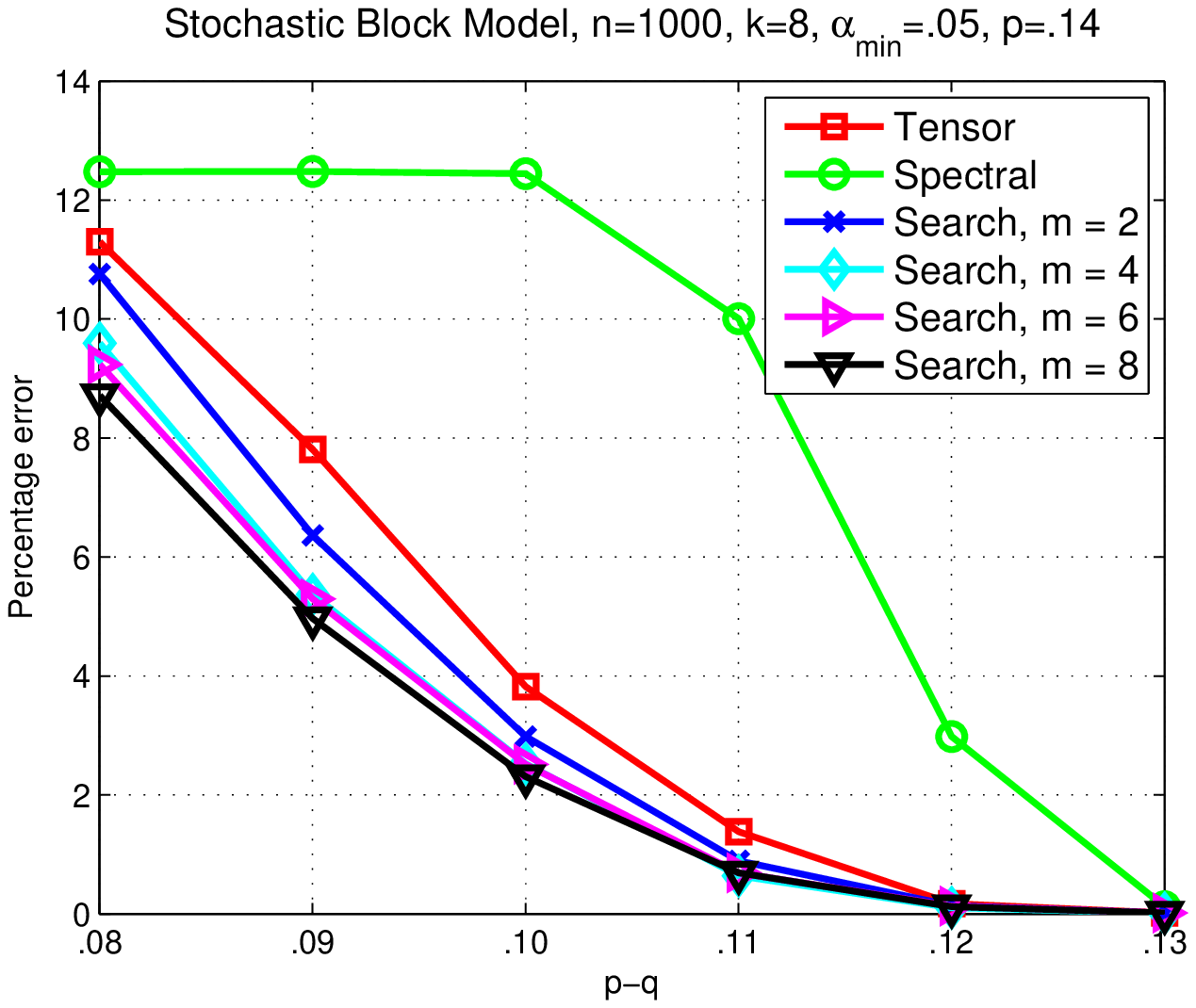}}
\caption{Labeled Nodes $+$ Parallel Clustering: Comparing the
  average error performance of Community Search algorithm with
  Spectral clustering \cite{NgJorWei02} and Tensor decomposition
  \cite{AndGeHsuKak13} algorithms in a stochastic block model with
  $n=1000,$ $k=8,$ $\alpha_{min}=.08,$ (a) $q=.01,$ (b) $p=.14.$ We consider the
  semi-supervised graph clustering setting when side information is
  available in the form of $m \in \{2,4,6,8\}$ labeled node from each
  community. The Community Search algorithm has a better performance
  over both Spectral clustering and Tensor
  decomposition.}
\label{fig:complexityLabeledParallel}
\end{figure}

Finally, we consider the semi-supervised graph clustering setting
described in Section \ref{sec:parallel_clustering} where we are
provided with $m$ labeled nodes from each community, and we want to
recover all communities. Recall that the Community Search algorithm
can also be used as a semi-supervised parallel graph clustering
algorithm. Figure \ref{fig:complexityLabeledParallel} plots the cumulative error over all
communities with increasing $p-q$ in a SBM with $n=1000,$ $k=8,$ and
using different number of labeled nodes. The weights in this case are
computed using the tree method and with radius $r=1.$ The Community
Search algorithm outperforms both Spectral clustering and Tensor
decomposition algorithms in both the experiments.

\subsection{Results on real dataset}

In this section we evaluate the performance of Community Search
algorithm on two real world networks.

In the first experiment we consider the
{\em US political blogosphere network} first introduced by
\cite{AdaGla:05blogosphere} where nodes correspond to political blogs
classified as either liberal or conservative during 2004 US election,
and edges represent hyperlinks between them. We consider the largest
connected component of the network having $1222$ nodes and $16,716$
edges. This dataset provides the ground-truth labels (liberal or
conservative) for each node; these labels were manually generated by authors
in \cite{AdaGla:05blogosphere} according to their content. The largest
component has two communities of sizes $586$ and $636$ according to
this ground-truth. 

\begin{table}[ht]
\caption{Average and best classification error (number of nodes that
  are misclassified in each estimated community compared to ground-truth)
  obtained by Community Search algorithm (W) with 
  Spectral clustering (S) \cite{NgJorWei02}  and Tensor decomposition (T)
  \cite{AndGeHsuKak13} algorithms on US political blogosphere
  network \cite{AdaGla:05blogosphere}. The Community Search
  algorithm achieves the best classification error of $53$ and better
  average error over the competing algorithms.
\label{tab:political_blogs}}
\begin{center}
\begin{tabular}{|c|c|c|c|c|c|c|c|c|}
  \hline 
    & W ($m=2$) & W ($m=4$) & W ($m=6$) & W ($m=8$) & W ($m=10$) & T & S\\
  \hline\hline
	 Mean & $56$ & $55.64$ & $55.32$ & $55.30$ & $54.98$ & $60$ & $70$ \\
  \hline
	 Best & $55$ & $54$ & $54$ & $53$ & $53$ & $60$ & $70$ \\
  \hline
  \hline
\end{tabular}
\end{center}
\end{table}

In this semi-supervised graph clustering setting, we randomly choose
$m \in \{2,4,6,8,10\}$ labeled nodes from each ground-truth community
as side information. Our performance metric is the classification
error, namely, the number of nodes wrongly classified in each estimated
community compared to the ground-truth communities\footnote{Since
  there are only two communities, the estimation error in the first
  community is equal to that in the second; thus, we can count any one of
  them.} ($e=|\{j \in V: j \in V_1, j \not \in \hat{V}_1 \ or \ j \in
\hat{V}_1, j \not \in V_1\}|$).

For each $m$ we observe the overall performance of the
Community Search algorithm over $50$ different random choices of
labeled nodes. As before, we compare the performance with Tensor
decomposition and Spectral clustering algorithms. For the Community
Search algorithm we compute weights using the tree method with radius
$r=1.$ In Table \ref{tab:political_blogs}, we show the best and average
classification error obtained by the clustering algorithms. With $r=1$
the Community Search algorithm shows a better classification error
than both Tensor and Spectral algorithms. In fact our algorithm
achieves the best classification error of $53,$ which is better than
other state--of--the--art algorithms
\cite{Jin:15fast,GaoMaZhangZhou:15} which achieved errors in
the range $58-60$ on this dataset. We also perform an in-depth analysis of the error cases in this dataset. We observed that $50$ nodes in the graph do not satisfy the community definition since they share fewer neighbors in their ground truth community (in degree) than the second community (out degree). Since the best error in our algorithm is $53,$ it appears that our algorithm performs close to the best achievable error in this dataset.

In our next experiment we consider the {\em Facebook--ego network} dataset from \cite{SnapData}, first introduced in \cite{LesMcau:12ego}. The network corresponds to a Facebook graph with $4039$ user nodes and $88,234$ edges. The dataset also contains $193$ ground-truth ego circles, where each circle is a group of user sharing a particular interest e.g. circle of college friends, family etc. Hence each circle here corresponds to a community. We consider the $10$ largest circles with more than $100$ nodes each as the ground-truth communities. Note that in this dataset some user can belong to multiple circles/communities, unlike a typical SBM. We remove such nodes from the ground-truth communities, but not from the graph. Even after this pruning step $9$ out of $10$ ground-truth communities have more than $100$ users each. Next, we try to recover these $9$ largest ground-truth communities by randomly choosing $m \in \{5,10,15\}$ nodes as labeled nodes. We run our Community
Search algorithm by computing weights using the tree method with radius $r=1,$ and we compare the average percentage estimation error (according to ground-truth) with that of Spectral clustering and Tensor decomposition algorithms. From the results presented in Table \ref{tab:facebook_ego} we observe that the Community Search algorithm has lower average error than the baselines even with just $m=5$ labeled nodes. Also, as expected the error reduces with increasing number of labeled nodes.

\begin{table}[ht]
\caption{Average percentage error obtained by Community Search algorithm (W) with 
  Spectral clustering (S) \cite{NgJorWei02}  and Tensor decomposition (T)
  \cite{AndGeHsuKak13} algorithms to estimate $9$ largest ego circles in Facebook--ego network \cite{LesMcau:12ego}. The Community Search
  algorithm achieves lower average error than the competing algorithms.
\label{tab:facebook_ego}}
\begin{center}
\begin{tabular}{|c|c|c|c|c|c|c|}
  \hline 
    & W ($m=5$) & W ($m=10$) & W ($m=15$) & T & S\\
  \hline\hline
	 Average Error & $4.204\%$ & $2.712\%$ & $2.708\%$ & $4.927\%$ & $4.462\%$ \\
  \hline
  \hline
\end{tabular}
\end{center}
\end{table}

\section{Conclusion and Discussion}

In this paper we defined the search problem in community detection, provided a simple generic framework for incorporating side information, and a corresponding algorithm to solve the problem. Our algorithm analytically matches the state of the art performance of existing algorithms that do not use side information, and empirically outperforms them on reliability and speed.

More generally, we believe that incorporating side information into graph analysis is a fertile and important area of research, as no real-world problem is a ``pure" graph problem (i.e. where the only input is a graph) of the kind studied in e.g. the vast majority of clustering literature.

There are several possible future directions: {\em (A)} Understanding fundamental limits of community detection \cite{MosNeemanSly14,Montanari15} when there is non-trivial side information (e.g. $\Theta(\log n)$ of labeled nodes in a community).  {\em (B)} Richer notions of side information, and corresponding problem definitions beyond search.  {\em (C)} From a more practical viewpoint we show in our experimental results (Section \ref{sec:experiments}) that even this simple form of side information can dramatically reduce the computation time for searching communities, and also improve error performance. As discussed in the previous section this work also provides a new method to parallelize graph clustering, an inherently difficult task. Adapting even faster algorithms, e.g. those based on belief-propagation, to this new semi-supervised setting is also an important prospect.

\section*{Acknowledgement}
We would like to acknowledge support from NSF grants
  CNS-1320175, 0954059, ARO grants W911NF-15-1-0227, W911NF-14-1-0387,
  W911NF-16-1-0377, and the US DoT supported D- STOP Tier 1 University Transportation Center. The authors also thank the Texas Advanced Computing Center \cite{TACC} at The University of Texas at Austin for providing HPC resources that have contributed to the research results reported within this paper.

%
\bibliographystyle{abbrvnat}
\bibliography{community_bibliography}  
%
%
\appendix
\section{Community Search: Proofs} \label{app:whiten_proof}

In this section we provide the proof details of Theorem \ref{thm:whiten_oracle}, Theorem \ref{thm:labeled_nodes} and the relevant Lemmas. 

\subsection{Community Search: Perturbation Analysis} \label{app:whiten_perturbation}

Let the expectation of the estimates $\hat{m}_1,$ $\hat{A}_1,$ $\hat{A}_2$ and $\hat{B}$ be represented by $m_1,$ $A_1,$ $A_2,$ $B$ respectively. Let $n_i=|P_i|$ be the size of partition $P_i.$ For a matrix $M,$ $\|M\|$ denotes its spectral norm. Recall that, 

\begin{eqnarray*}
A_1 &=& \frac{1}{\sqrt{n_3}} E[X_{P_1,P_3}] \ , \ A_2 = \frac{1}{\sqrt{n_3}} E[X_{P_2,P_3}] \\
m_1 &=& \sum_{i=1}^k \alpha_i \mu_{P_1,i} \ , \ B = \sum_{i=1}^k \alpha_i \omega_i \mu_{P_2,i}^T 
\end{eqnarray*}

where $\omega_i=E[w_j | j \in V_i].$ Let the rank $k$-svd of $A_1,A_2$ be given by $A_1 = U_1 D_1 V_1^T,$ $A_2 = U_2 D_2 V_2^T,$ and for the estimates $\hat{A}_1 = \hat{U}_1 \hat{D}_1 \hat{V}_1^T,$ $\hat{A}_2 = \hat{U}_2 \hat{D}_2 \hat{V}_2^T.$

\begin{lemma} \label{lem:hatWAhatWHalf_ub}
Let $\max \{\|\hat{A}_1-A_1\|,\|\hat{A}_2-A_2\|\} \leq \epsilon_2$ and $\epsilon_2 < \min\{\sigma_k(A_1),\sigma_k(A_2)\}/12.$ Let $\hat{W}_1=\hat{U}_1\hat{D}_1^{-1},\hat{W}_2=\hat{U}_2\hat{D}_2^{-1}$ be the whitening matrices. Then,

\begin{eqnarray*}
\|I_k-(\hat{W}_1^TA_1A_1^T\hat{W}_1)^{1/2}\| &\leq& \frac{6 \epsilon_2}{\sigma_k(A_1)}\\
\|I_k-(\hat{W}_1^TA_1A_1^T\hat{W}_1)^{-1/2}\| &\leq& \frac{12 \epsilon_2}{\sigma_k(A_1)}\\
\|I_k-(\hat{W}_2^TA_2A_2^T\hat{W}_2)^{-1/2}\| &\leq& \frac{12 \epsilon_2}{\sigma_k(A_2)}
\end{eqnarray*}

\end{lemma}
\begin{proof}
We prove this along the lines in \cite{HsuKakade13}. The matrix $\hat{W}_1$ whitens $\hat{A}_1\hat{A}_1^T$ since,

$$
\hat{W}_1^T \hat{A}_1\hat{A}_1^T \hat{W}_1 = \hat{D}_1^{-1}\hat{U}_1^T\hat{A}_1\hat{A}_1^T\hat{U}_1\hat{D}_1^{-1}=I_k
$$

Similarly $\hat{W}_2$ whitens $\hat{A}_2\hat{A}_2^T.$ 

Also note $\epsilon_2<\sigma_k(A_1)/2,$ hence using Weyl's inequality $\sigma_k(\hat{A}_1)\geq \sigma_k(A_1)/2.$ This implies

\begin{eqnarray*}
\|I_k-\hat{W}_1^TA_1A_1^T\hat{W}_1\| &=& \|\hat{W}_1^T(\hat{A}_1\hat{A}_1^T-A_1A_1^T)\hat{W}_1\| \\
&\leq& \|\hat{W}_1^T\hat{A}_1(\hat{A}_1^T-A_1^T)\hat{W}_1\|+\|\hat{W}_1^T(\hat{A}_1-A_1)A_1^T\hat{W}_1\|  \\
&\leq& \|\hat{W}_1^T\hat{A}_1\| \|\hat{A}_1^T-A_1^T\|\|\hat{W}_1\| + \|\hat{W}_1^T\| \|\hat{A}_1-A_1\| \|A_1^T\hat{W}_1\| \\
&<& \frac{2 \epsilon_2}{\sigma_k(A_1)}+\frac{2 \epsilon_2}{\sigma_k(A_1)}\left(\|\hat{A}_1^T\hat{W}_1\|+\|(A_1^T-\hat{A}_1^T)\hat{W}_1\|\right) \\
&<& \frac{2 \epsilon_2}{\sigma_k(A_1)}+\frac{2 \epsilon_2}{\sigma_k(A_1)}\left(1+\frac{2 \epsilon_2}{\sigma_k(A_1)}\right) \\
&\leq& \frac{6 \epsilon_2}{\sigma_k(A_1)}
\end{eqnarray*}

Therefore all eigenvalues of the matrix $\hat{W}_1^TA_1A_1^T\hat{W_1}$ lie in the interval $\left(1-\frac{6 \epsilon_2}{\sigma_k(A_1)},1+\frac{6 \epsilon_2}{\sigma_k(A_1)}\right).$ This implies the eigenvalues of $(\hat{W}_1^TA_1A_1^T\hat{W}_1)^{1/2}$ also lie in the same interval and that of  $(\hat{W}_1^TA_1A_1^T\hat{W}_1)^{-1}$ lie in the interval $\left(1/(1+6 \epsilon_2/\sigma_k(A_1)),1/(1-6 \epsilon_2/\sigma_k(A_1)\right)).$ The first bound follows directly. To show the second bound we compute,

\begin{eqnarray*}
(I_k-(\hat{W}_1^TA_1A_1^T\hat{W}_1)^{-1/2})(I_k+(\hat{W}_1^TA_1A_1^T\hat{W}_1)^{-1/2}) &=& I_k-(\hat{W}_1^TA_1A_1^T\hat{W}_1)^{-1} \\
I_k-(\hat{W}_1^TA_1A_1^T\hat{W}_1)^{-1/2} &=& \left(I_k-(\hat{W}_1^TA_1A_1^T\hat{W}_1)^{-1}\right)\times \\
&& (I_k+(\hat{W}_1^TA_1A_1^T\hat{W}_1)^{-1/2})^{-1} \\
\|I_k-(\hat{W}_1^TA_1A_1^T\hat{W}_1)^{-1/2}\| &\leq& \|I_k-(\hat{W}_1^TA_1A_1^T\hat{W}_1)^{-1}\| \\
&\leq& \frac{1}{1-6 \epsilon_2/\sigma_k(A_1)}-1 \\
&\leq& \frac{12 \epsilon_2}{\sigma_k(A_1)}
\end{eqnarray*}

Similarly we can show the second bound using $\hat{W}_2$ and $A_2.$
\end{proof}

\begin{lemma} \label{lem:uhat_ub}
Let $\|\hat{R}-R\|\leq \delta < \sigma_2(R)/2.$ $u_1$ be a left singular vector of $R$ corresponding to the largest singular value and $\hat{u}_1$ be that of $\hat{R}.$ Then,

$$
\|\hat{u}_1-u_1\| \leq \frac{8\delta}{(\sigma_1(R)-\sigma_2(R))}
$$
\end{lemma}
\begin{proof}
The result follows from the generalized sin--$\theta$ theorem by \cite{wedin:72}. In particular we use an useful version of it from \cite{Yu:15SinTheta} [Theorem $4$]. We get,

\begin{eqnarray*}
\|\hat{u}_1-u_1\| &\leq& \frac{2^{3/2}(2\sigma_1(R)+\|\hat{R}-R\|)\|\hat{R}-R\|}{\sigma_1(R)^2-\sigma_2(R)^2} \\
&\leq& \frac{2^{3/2}(2\sigma_1(R)+2\sigma_2(R))\|\hat{R}-R\|}{(\sigma_1(R)-\sigma_2(R))(\sigma_1(R)+\sigma_2(R))} \\
&\leq& \frac{8\delta}{(\sigma_1(R)-\sigma_2(R))}
\end{eqnarray*}
\end{proof}

\begin{lemma} \label{lem:zhat_ub}
Assume $\|\hat{u}_1-u_1\| \leq \eta_1,$ $\|\hat{A}_1-A_1\| \leq \epsilon_2.$ Let $\hat{z}=\hat{U}_1 \hat{D}_1 \hat{u}_1.$ $z$ be given by the equation $u_1=W_1^Tz,$ where $W_1=\hat{W}_1(\hat{W}_1^TA_1A_1^T\hat{W}_1)^{-1/2}.$ Then,

\begin{equation*}
\|\hat{z}-z\| \leq 2\sigma_1(A_1)\eta_1 + \frac{16\sigma_1(A_1)\epsilon_2}{\sigma_k(A_1)}
\end{equation*}

\end{lemma}
\begin{proof}
First using Wedin's theorem \cite{wedin:72} we get,

\begin{equation}
\|\hat{U}_1\hat{U}_1^T-U_1U_1^T\| \leq \frac{4\epsilon_2}{\sigma_k(A_1)} \label{eq:wedin}
\end{equation}

We can bound $\hat{z}-z$ as follows.

\begin{eqnarray}
\|\hat{z}-z\| &=& \|\hat{z}-U_1U_1^Tz\| \leq \|\hat{z}-\hat{U}_1\hat{U}_1^Tz\|+\|\hat{U}_1\hat{U}_1^T-U_1U_1^T\| \|z\| \nonumber \\
&\stackrel{(a)}{\leq}& \|\hat{z}-\hat{U}_1\hat{U}_1^Tz\|+\frac{4\epsilon_2 \|z\|}{\sigma_k(A_1)} \nonumber \\
&\stackrel{(b)}{=}& \|\hat{z}-\hat{U}_1\hat{U}_1^Tz\|+\frac{4\epsilon_2 \|U_1D_1u'\|}{\sigma_k(A_1)} \\
&\leq& \|\hat{z}-\hat{U}_1\hat{U}_1^Tz\|+\frac{4\sigma_1(A_1)\epsilon_2}{\sigma_k(A_1)} \label{eq:zhat_1} 
\end{eqnarray}

The step (a) uses equation \ref{eq:wedin}, and step (b) uses the fact that the matrix $D_1^{-1}U_1^T$ also whitens $A_1A_1^T,$ therefore $z$ can also be expressed as $z=U_1D_1u'$ for some unit vector $u'.$ Since $u_1=W_1^Tz,$ we can write also $\hat{D}_1(\hat{W}_1^TA_1A_1^T\hat{W}_1)^{1/2}u_1=\hat{U}_1^Tz.$ Now we bound the first term.

\begin{eqnarray}
\|\hat{z}-\hat{U}_1\hat{U}_1^Tz\| &=& \|\hat{U}_1\hat{D}_1\hat{u}_1-\hat{U}_1\hat{U}_1^Tz\| \nonumber \\
&=&\|\hat{U}_1\hat{D}_1(\hat{u}_1-u_1)+\hat{U}_1\hat{D}_1u_1-\hat{U}_1\hat{U}_1^Tz\| \nonumber \\
&\leq& \|\hat{D}_1\|\|\hat{u}_1-u_1\|+\|\hat{U}_1\hat{D}_1u_1-\hat{U}_1\hat{U}_1z\|  \nonumber \\
&\leq& 2\sigma_1(A_1)\eta_1 + \|\hat{U}_1\hat{D}_1u_1-\hat{U}_1\hat{D}_1(\hat{W}_1^TA_1A_1^T\hat{W}_1)^{1/2}u_1\| \nonumber \\
&\leq& 2\sigma_1(A_1)\eta_1 + \|\hat{U}_1\hat{D}_1(I_k-(\hat{W}_1^TA_1A_1^T\hat{W}_1)^{1/2})\|\|u_1\|\nonumber \\
&\leq& 2\sigma_1(A_1)\eta_1 + \|\hat{D}_1\|\|I_k-(\hat{W}_1^TA_1A_1^T\hat{W}_1)^{1/2}\| \nonumber \\
&\leq& 2\sigma_1(A_1)\eta_1 + \frac{12\sigma_1(A_1)\epsilon_2}{\sigma_k(A_1)} 
\end{eqnarray}

where the last inequality follows from Lemma \ref{lem:hatWAhatWHalf_ub}. Combining the above with equation \ref{eq:zhat_1} we get,

$$
\|\hat{z}-z\| \leq 2\sigma_1(A_1)\eta_1 + \frac{12\sigma_1(A_1)\epsilon_2}{\sigma_k(A_1)}+\frac{4\sigma_1(A_1)\epsilon_2 }{\sigma_k(A_1)} = 2\sigma_1(A_1)\eta_1 + \frac{16\sigma_1(A_1)\epsilon_2}{\sigma_k(A_1)}
$$
\end{proof}

\subsection{Community Search: Concentration} \label{sec:whiten_concentration}

In this section using concentration bounds we compute the parameter range of $p,q,k,\alpha_i$ for which the Community Search algorithm can recover the particular community membership vector $\mu_1$ with high probability. For ease of exposition for this section we assume the partitions $P_1,P_2,P_3,P_4$ are of equal size. Therefore $n_1=n_2=n_3=n_4 = \frac{n}{4}.$ However the results easily generalize to any random unbiased split. Now we restate the Matrix Bernstein inequality \cite{Tropp:15} and then use it to bound the perturbation of the estimates $\hat{A}_1, \hat{A}_2.$

\begin{theorem}[Matrix Bernstein] \label{thm:matrix_bernstein}
Let $\{A_j\}_{j=1}^n$ be a sequence of i.i.d. real random $d_1 \times d_2$ matrices such that $E[A_j]=0,$ $\|A_j\| \leq L.$ Define $Z=\sum_{j=1}^n A_j.$ Let $\sigma^2 = \max \{\|E[Z Z^T]\|,\|E[Z^T Z]\|\}.$ Then for all $t \geq 0,$

\begin{equation*}
P\left( \|Z\| \geq t\right) \leq (d_1+d_2)\exp \left(\frac{-t^2/2}{\sigma^2+Lt/3}\right)
\end{equation*}
\end{theorem}

\begin{lemma}[Concentration of $\hat{A}_1,\hat{A}_2$] \label{lem:A1hat_ub}
Let $\hat{A}_1, \hat{A}_2$ be as given in Algorithm \ref{alg:comm_search_white}. Then,

\begin{eqnarray*}
\|\hat{A}_1-A_1\| &=& O\left(\sqrt{p\log \frac{(n_1+n_3)}{\delta}}\right) \\
\|\hat{A}_2-A_2\| &=& O\left(\sqrt{p\log \frac{(n_2+n_3)}{\delta}}\right) \\
\end{eqnarray*}

with probability greater than $1-2\delta.$
\end{lemma}
\begin{proof}
Note that we can write $\hat{A}_1 = \frac{1}{\sqrt{n_3}} \sum_{j \in P_3} X_{P_1,j}e_j^T,$ where $e_j$ is the unit vector with $1$ in the $j$-th coordinate. Then $Z=\hat{A}_1-A_1 = \frac{1}{\sqrt{n_3}} \sum_{j \in P_3} (X_{P_1,j}-\mu_{P_1,c_j})e_j^T=\sum_{j \in P_3} Z_j,$ $c_j$ being the cluster of $j$-th node. Then,

\begin{eqnarray*}
\|E[ZZ^T]\| &=& \|\sum_{j \in P_3} E[Z_jZ_i^T]\| \leq \sum_{j \in P_3} \|E[Z_j Z_j^T]\| \\ 
&=& \sum_{j \in P_3} \frac{1}{n_3} \|E[(X_{P_1,j}-\mu_{P_1,c_j})(X_{P_1,j}-\mu_{P_1,c_j})^T]\| \\
&\leq& \sum_{j \in P_3} \frac{1}{n_3} p(1-p) \leq p
\end{eqnarray*}
 
Also,

\begin{equation*}
\|E[Z^T Z]\| = \|\sum_{j \in P_3} E[Z_j^T Z_j]\|= \|\sum_{j \in P_3} \frac{1}{n_3} E[\|X_{P_1,j}-\mu_{P_1,c_j}\|^2 e_j e_j^T ]\|\leq \frac{n_1 p}{n_3}
\end{equation*}

Assuming $n_1=n_3$ we have the variance term bounded by $\sigma^2 = \max\{\|E[ZZ^T]\|,\|E[Z Z^T]\|\} = p.$ Now with high probability $\|Z_j\|=\frac{1}{n_3}\|(X_{P_1,j}-\mu_{P_1,c_j})e_j^T\| \leq \sqrt{2p}:=L.$ Therefore by applying from Matrix-Bernstein inequality with probability greater than $1-\delta$,

$$
\|\hat{A}_1-A_1\| \leq 2\sigma \sqrt{\log \frac{(n_1+n_3)}{\delta}} = O \left( \sqrt{p\log \frac{(n_1+n_3)}{\delta}} \right)
$$

Similarly we find the second bound for $\|\hat{A}_2-A_2\|.$
\end{proof}

\begin{lemma}[Whitening matrix concentration] \label{lem:W_conc}
Assume that $\max\{\|\hat{A}_1-A_1\|,\|\hat{A}_2-A_2\|\}<\epsilon_2,$ and $\epsilon_2<\min\{\sigma_k(A_1),\sigma_k(A_2)\}/4.$ Let $\hat{W}_1=\hat{U}_1\hat{D}_1^{-1},\hat{W}_2=\hat{U}_2\hat{D}_2^{-1}$ be the whitening matrices. Define $W_1:=\hat{W}_1(\hat{W}_1^TA_1A_1^T\hat{W}_1)^{-1/2}$ and $W_2:=\hat{W}_2(\hat{W}_2^TA_2A_2^T\hat{W}_2)^{-1/2}.$ Then,

\begin{eqnarray*}
\|\hat{W}_1-W_1\| &=& O\left(\frac{\sqrt{p \log n_1}}{\alpha_{min}^2n_1(p-q)^2}\right) \\
\|\hat{W}_2-W_2\| &=& O\left(\frac{\sqrt{p \log n_2}}{\alpha_{min}^2 n_2(p-q)^2}\right)
\end{eqnarray*}

\end{lemma}
\begin{proof}
First note that the matrix $W_1$ whitens the matrix $A_1A_1^T$ since,

$$
W_1^TA_1A_1^TW_1 = (\hat{W}_1^TA_1A_1^T\hat{W}_1)^{-1/2}\hat{W}_1^TAA^T\hat{W}_1 (\hat{W}_1^TA_1A_1^T\hat{W}_1)^{-1/2} = I_k
$$

Similarly $W_2$ whitens matrix $A_2A_2^T.$ We can bound the perturbation as follows.

\begin{eqnarray*}
\|\hat{W}_1-W_1\| &=& \|\hat{W}_1(I_k-(\hat{W}_1^TA_1A_1^T\hat{W})_1^{-1/2})\| \\
&\leq& \|\hat{W}_1\| \|I_k-(\hat{W}_1^TA_1A_1^T\hat{W}_1)^{-1/2}\| \\
&\leq& \frac{2}{\sigma_k(A)}\times \frac{12\epsilon_2}{\sigma_k(A_1)}=\frac{24 \epsilon_2}{\sigma_k(A_1)^2}
\end{eqnarray*}

where the last inequality follows from Lemma \ref{lem:hatWAhatWHalf_ub}. Now from Lemma \ref{lem:A1hat_ub} we have $\epsilon_2=O(\sqrt{p \log n_1}).$ Also observe that  $\sigma_k(A_1)=\Omega(\alpha_{min}\sqrt{n_1}(p-q)).$ Using these in the above bound we get

$$
\|\hat{W}_1-W_1\| = O\left(\frac{\sqrt{p \log n_1}}{\alpha_{min}^2n_1(p-q)^2}\right)
$$

The second bound for $\|\hat{W}_2-W_2\|$ follows.
\end{proof}

\begin{lemma}[$R$ matrix concentration] \label{lem:WBW_ub}
Let $\hat{R}=\hat{W}_1^T \hat{B} \hat{W}_2$ and $R=W_1^T B W_2$ then,

$$
\|\hat{R}-R\| = \tilde{O}\left(\max\left\{\frac{\alpha_{max}^2p^{2.5}\gamma_1}{\alpha_{min}^3\sqrt{n}(p-q)^3},\frac{\alpha_{max}^2 p^2 \gamma_2}{\alpha_{min}^2 (p-q)^2}\right\} \right)
$$

where $\gamma_1=\max_{j\in P_4} |\hat{w}_j|,$ and $\gamma_2 = \max_{j \in P_4} |\hat{w}_j-\bar{w}_j|.$
\end{lemma}
\begin{proof}
Let $c_j \in [k]$ denote the community for the $j$-th node. We can upper bound the estimation error in $R$ matrix as follows.

\begin{eqnarray*}
\|\hat{R}-R\| &=& \|\hat{W}_1^T \hat{B} \hat{W}_2 - W_1^T B W_2\| = \frac{1}{n_4} \|\sum_{j \in P_4} (\hat{w}_j \hat{W}_1^T X_{P_1,j} X_{P_2,j}^T \hat{W}_2 - \bar{w}_j W_1^T \mu_{P_1,c_j} \mu_{P_2,c_j}^T W_2)\| \\
&\leq& T_1 + T_2 + T_3 + T_4 + T_5
\end{eqnarray*}

where,

\begin{eqnarray*}
T_1 &=& \|\frac{1}{n_4} \sum_{j \in P_4} \hat{w}_j \hat{W}_1^T (X_{P_1,j}-\mu_{P_1,c_j})X_{P_2,j}^T \hat{W}_2\| = \|\hat{W}_1^T (\hat{A}_1-A_1) diag(\hat{w}_1 , ... ,\hat{w}_{n_4}) \hat{A}_2^T \hat{W}_2\|\\
T_2 &=& \|\frac{1}{n_4} \sum_{j \in P_4} \hat{w}_j \hat{W}_1^T \mu_{P_1,c_j} (X_{P_2,j}-\mu_{P_2,c_j})^T \hat{W}_2\| = \|\hat{W}_1^T A_1 diag(\hat{w}_1 , ... ,\hat{w}_{n_4}) (\hat{A}_2 - A_2)^T \hat{W}_2\|\\
T_3 &=& \|\frac{1}{n_4} \sum_{j \in P_4} \hat{w}_j (\hat{W}_1-W_1)^T \mu_{P_1,c_j} \mu_{P_2,c_j}^T \hat{W}_2\| = \|(\hat{W}_1-W_1)^T A_1 diag(\hat{w}_1 , ... ,\hat{w}_{n_4}) A_2^T \hat{W}_2\|\\
T_4 &=& \|\frac{1}{n_4} \sum_{j \in P_4} \hat{w}_j W_1 \mu_{P_1,c_j} \mu_{P_2,c_j}^T (\hat{W}_2-W_2)\| = \|W_1^T A_1 diag(\hat{w}_1 , ... ,\hat{w}_{n_4}) A_2^T (\hat{W}_2-W_2)\|\\
T_5 &=& \|\frac{1}{n_4} \sum_{j \in P_4} (\hat{w}_j-\bar{w}_j) W_1 \mu_{P_1,c_j} \mu_{P_2,c_j}^T W_2\| = \|W_1^T A_1 diag(\hat{w}_1-\bar{w}_1 , ... ,\hat{w}_{n_4}-\bar{w}_{n_4}) A_2^T W_2\|
\end{eqnarray*}

Let $\gamma_1=\max_{j\in P_4} |\hat{w}_j|,$ and $\gamma_2 = \max_{j \in P_4} |\hat{w}_j-\bar{w}_j|.$ Then using Lemmas \ref{lem:A1hat_ub} and \ref{lem:W_conc} we get $T_1=T_2=\tilde{O}\left(\frac{\alpha_{max}p^{1.5}\gamma_1}{\alpha_{min}^2\sqrt{n}(p-q)^2} \right),$  $T_3=T_4=\tilde{O}\left(\frac{\alpha_{max}^2p^{2.5}\gamma_1}{\alpha_{min}^3\sqrt{n}(p-q)^3} \right),$ and $T_5=\tilde{O}\left(\frac{\alpha_{max}^2 p^2 \gamma_2}{\alpha_{min}^2 (p-q)^2} \right).$ The dominating term is given by the maximum of $T_3,T_4$ and $T_5.$
\end{proof}

\begin{lemma}[Thresholding] \label{lem:threshold}
Let $e=\hat{z}-z,$ threshold $\tau=\sqrt{\alpha_1}(p+q)/2.$ Let $E=\{i \in V: i \in \hat{V}_1, i \not \in V_1 \ or \ i \in V_1, i \not \in \hat{V}_1\}$ be the set of erroneous nodes after thresholding. If $\|e\| = o\left(\alpha_1 \sqrt{n}(p-q)\right)$ then $|E|=o(\alpha_1 n).$   
\end{lemma}
\begin{proof}
We prove this along similar lines in \cite{AndGeHsuKak13}. Note that $z=\sqrt{\alpha_1} \mu_1$ is a vector with all coordinates either $\sqrt{\alpha_1} p$ or $\sqrt{\alpha_1}q.$ Since the threshold is $\tau=\sqrt{\alpha_1} (p+q)/2,$ this implies error in any node $i \in E$ is caused when the magnitude error in the corresponding coordinate $\hat{z}_{i}$ is at least $\sqrt{\alpha_1}(p-q)/2.$ Let $e_i$ denote the magnitude error in the $i$-th coordinate. Then,
\begin{eqnarray*}
\|e\|^2 &=& \sum_{i \in E} e_i^2 + \sum_{i \in V\backslash E} e_i^2 \\
\|e\|^2 &\geq& \sum_{i \in E} e_i^2 \geq |E| \alpha_1 (p-q)^2/4
\end{eqnarray*} 

Now since $\|e\|$ is upper bounded as $\|e\| = o\left(\alpha_1 \sqrt{n}(p-q)\right)$ then it follows that the number of error is bounded by $|E| = o(\alpha_1 n).$
\end{proof}

{\bf Proof of Theorem \ref{thm:whiten_oracle}:}
\begin{proof}
We observe that the vector $z$ in Algorithm \ref{alg:whiten_subroutine} is simply a scalar multiple of the partial community membership vector estimate $\hat{\mu}_{P_1}.$ Therefore we can also recover community $1$ subset $V_{P_1}$ by directly thresholding $z$ using a threshold $\tau' = \sqrt{\alpha_1}(p+q)/2.$ In other words the threshold $\tau$ required in Algorithm \ref{alg:comm_search_white} is simply $\tau'/a,$ $a$ as defined in Algorithm \ref{alg:whiten_subroutine}. Therefore it is sufficient show that the estimated vector $\hat{z}$ is close enough to the true vector $z$ and use Lemma \ref{lem:threshold} to guarantee we can estimate the community with only $o(1)$ fraction error. 

In Lemma \ref{lem:WBW_ub} note that the maximum weight $\gamma_1 \leq \sigma_1(R)+\gamma_2.$ Then using conditions (A2) and (A3) in Lemma \ref{lem:WBW_ub} we get with high probability,

\begin{equation}
\|\hat{R}-R\| \leq C_1 (\sigma_1(R)-\sigma_2(R)) \frac{\alpha_{min}^2 (p-q)^2}{\alpha_{max}^2 p^2 \xi(n)} \label{eq:Rhat_ub2}
\end{equation}

for some constant $C_1.$ Therefore from Lemma \ref{lem:uhat_ub} and equation \ref{eq:Rhat_ub2} we get,

\begin{equation}
\|\hat{u}-u\| \leq \frac{8\|\hat{R}-R\|}{(\sigma_1(R)-\sigma_2(R))}=\leq 8C_1 \frac{\alpha_{min}^2 (p-q)^2}{\alpha_{max}^2 p^2 \xi(n)}:=\eta_1
\end{equation}

From Lemma \ref{lem:A1hat_ub} we get $\epsilon_2=O(\sqrt{p \log n}).$ Also note that $\|z\|=O(\sqrt{\alpha_1n}p),$ $\sigma_1(A_1)=O(\alpha_{max}\sqrt{n}p),$ and $\sigma_k(A_1)=\Omega(\alpha_{min}\sqrt{n}(p-q)).$ Now using the above bound $\eta_1$ in Lemma \ref{lem:zhat_ub} we bound $\|\hat{z}-z\|$ as follows. 

\begin{eqnarray}
\|\hat{z}-z\| &\leq& 2\sigma_1(A_1)\eta_1 + \frac{16\sigma_1(A_1)\epsilon_2}{\sigma_k(A_1)} \nonumber \\
&\leq& 2\alpha_{max}\sqrt{n}p \times 8C_1\frac{\alpha_{min}^2 (p-q)^2}{\alpha_{max}^2 p^2 \xi(n)} + 16C_2 \frac{\alpha_{max}\sqrt{n} p\sqrt{p \log n}}{\alpha_{min}\sqrt{n}(p-q)} \nonumber \\
&=& 16C_1\frac{\alpha_{min}^2 \sqrt{n}(p-q)^2}{\alpha_{max} p \xi(n)} + 16C_2 \frac{\alpha_{max}p^{1.5}\sqrt{\log n}}{\alpha_{min}(p-q)} \nonumber \\
&\stackrel{(a)}{\leq}& C_3\alpha_{min} \frac{\sqrt{n} (p-q)}{\sqrt{\xi(n)}} \nonumber \\
&=& o\left( \alpha_{1} \sqrt{n} (p-q) \right) \label{eq:what_con_ub}
\end{eqnarray}

where $C_1,C_2,C_3$ are constants. Step (a) follows from condition (A3). Now applying Lemma \ref{lem:threshold} for the partition $P_1$ it follows that by thresholding $\hat{z}$ using threshold $\tau=\sqrt{\alpha_1}(p+q)/2$ the number of erroneous nodes in $\hat{V}_{P_1}$ is bounded as $|E|=o(\alpha_1 n).$ Similarly with high probability this holds for partitions $P_2,P_3$ and $P_4$ as well. Therefore we can recover community $V_1$ with $o(1)$ fraction error with high probability.
\end{proof}

{\bf Proof of Theorem \ref{thm:search_exact_recovery}:}
\begin{proof}
Under conditions (A1)-(A3) using Theorem \ref{thm:whiten_oracle} we can guarantee that the estimated community $\hat{V}_{P_1}$ has at most $o(\alpha_1 n)$ erroneous nodes. Now consider the following degree thresholding step. For any $j\in P_2$ $d_{j}(\hat{V}_{P_1})$ be the number of edges $j$ share with nodes in $\hat{V}_{P_1}.$ Define $\hat{V}_{P_2}:=\{j \in P_2 : d_{j}(\hat{V}_{P_1})\geq \tau''\},$ for $\tau''=|V_1 \cap P_1|(p+q)/2.$ Then we claim that $\hat{V}_{P_2}=V_1 \cap P_2$ with high probability. 

Note that the edges between $P_1$ and $P_2$ are not used in Algorithm \ref{alg:comm_search_white}. Let $v_1=|\hat{V}_{P_1} \cap V_1|$ be the number of correct nodes, and $e_1=|\hat{V}_{P_1} \cap V_1^c|$ be the number of erroneous nodes in $\hat{V}_{P_1}.$ Theorem \ref{thm:whiten_oracle} asserts with high probability $v_1=\Theta(\alpha_1 n),$ $e_1=o(\alpha_1 n).$ Let $v_0=|V_1 \cap P_1|=\Theta(\alpha_1 n).$ Note that $v_1 \geq v_0-e_1.$ Now for any $j \in V_1 \cap P_2$ using Chernoff bound we get with high probability

\begin{eqnarray*}
d_{j}(\hat{V}_{P_1}) &\geq& v_1 p - \sqrt{v_1 p \log n} + e_1 q - \sqrt{e_1 q \log n} \\
&\geq& v_0 p - e_1 p - \sqrt{v_1 p \log n} + e_1 q - \sqrt{e_1 q \log n} \\
&=& v_0 \frac{(p+q)}{2} + \left[v_0 \frac{(p-q)}{2} - e_1(p-q) - \sqrt{v_1 p \log n} - \sqrt{e_1 q \log n} \right] \\
&\geq& v_0 \frac{(p+q)}{2}
\end{eqnarray*} 

where the last step follows since $e_1=o(\alpha_1 n)$ and using condition (A3). Similarly for any node $j \in V_1^c \cap P_2$ using Chernoff bound we can get with high probability

\begin{eqnarray*}
d_{j}(\hat{V}_{P_1}) &\leq& v_1 q + \sqrt{v_1 q \log n} + e_1 p + \sqrt{e_1 p \log n} \\
&\leq& v_0 q + \sqrt{v_1 q \log n} + e_1 p + \sqrt{e_1 p \log n} \\
&=& v_0 \frac{(p+q)}{2} - \left[v_0 \frac{(p-q)}{2} -  \sqrt{v_1 q \log n} - e_1 p - \sqrt{e_1 p \log n}\right]\\
&<& v_0 \frac{(p+q)}{2}
\end{eqnarray*}

again using the fact $e_1=o(\alpha_1 n)$ and using condition (A3). Therefore using a threshold $v_0 \frac{(p+q)}{2}$ we can correctly recover all nodes in $V_1 \cap P_2.$ Now by rotating the partitions and repeatedly applying Algorithm \ref{alg:comm_search_white} $+$ degree thresholding we can correctly recover community $V_1$ with high probability. This concludes the proof. 
\end{proof}

\subsection{Recovery via Labeled Nodes} \label{app:labeled_node_proof}

{\bf Proof of Theorem \ref{thm:labeled_nodes}:}
\begin{proof}
First note that edges between partitions $P_1$ and $P_2$ are not used in Algorithm \ref{alg:comm_search_white}. Therefore the weights can be computed using these edges so that they are independent of the remaining algorithm. For this proof we assume $\mathcal{L}$ to be the set of labeled nodes in partition $P_2.$ For each node $i\in P_1$ we consider a tree of radius $r$ in this partition with $i$ as root, then count the number of edges from the leaves of this tree and labeled node set $\mathcal{L}$ in partition $P_2.$ Let $T_i(r)$ be the subgraph of all nodes at a distance less than or equal to $r$ from node $i.$ When $p=\Theta\left(\frac{\log n}{n^\epsilon}\right),$ $q=\Theta\left(\frac{\log n}{n^\epsilon}\right)$ applying Chernoff bound it is easy to see as long as $|T_i(r)| = o\left(\frac{n^{\epsilon}}{\log n} \right)$ then with high probability $T_i(r)$ is a tree. With $L=\Omega(n^{\epsilon/2}\sqrt{\log n}),$ and for $r \leq \frac{2\log(n^\epsilon /L)}{\log np}=\frac{2\epsilon \log n-2\log L}{(1-\epsilon)\log n + O(\log\log n)}$ this holds. Now starting from a node in community $i$ let $f_j^i(t)$ be the number of nodes in community $j$ at a distance $t$ from the root node. Since $f_j^i(t)<|T_i(r)|$ this implies $f_j^i(t)=o(\alpha n).$ Now consider the following set of $k$ recursive equations.
\begin{equation}
\bar{f}_j^i(t) = \alpha n p \bar{f}_j^i(t-1) + \sum_{l \neq j} \alpha n q \bar{f}_l^i(t-1) \label{eq:mean_recursive}
\end{equation}
for $j \in [k].$ Since the number of nodes in community $l$ at distance $t$ which are neighbors of $f_j^i(t-1)$ nodes in community $j$ at distance $t-1$ are binomially distributed with probability $p$ if $l=j$ or probability $q$ otherwise; we can use Chernoff bound to see that with high probability the actual number of nodes $f_j^i(t)$ can be expressed as
$$
f_j^i(t) = \bar{f}_j^i(t) + o(\bar{f}_j^i(t))
$$
Now the initial condition in the recursive equation \eqref{eq:mean_recursive} is given by $\bar{f}_{j}^i(0)=1$ when $j=i,$ $\bar{f}_{j}^i(0)=0$ otherwise. We will prove the theorem in three steps. 

{\bf Claim 1:} {\em $\bar{f}_i^i(t)>\bar{f}_j^i(t)$ for all $t$ and $j \neq i$}

We prove this by induction. From the initial conditions $\bar{f}_i^i(0)=1>0=\bar{f}_j^i(0),$ so the claim holds for $t=0.$ Assume it holds for $t-1.$ Then for all $j \neq i$,
\begin{eqnarray*}
\bar{f}_i^i(t) &=& \alpha n p \bar{f}_i^i(t-1) + \alpha n q \bar{f}_j^i(t-1) + \sum_{l \neq i,j} \alpha n q \bar{f}_l^i(t-1) \\
\bar{f}_j^i(t) &=& \alpha n q \bar{f}_i^i(t-1) + \alpha n p \bar{f}_j^i(t-1) + \sum_{l \neq i,j} \alpha n q \bar{f}_l^i(t-1) \\
\end{eqnarray*}
Then,
\begin{equation}
\bar{f}_i^i(t)-\bar{f}_j^i(t) = \alpha n (p-q) (\bar{f}_i^i(t-1)-\bar{f}_j^i(t-1)) > 0 \label{eq:fbart_gap}
\end{equation}
which follows from the induction hypothesis and since $p>q.$ This asserts that the claim is true.

{\bf Claim 2:} {\em $E[w_l|l \in V_1] > E[w_l | l \in V_i]$ for all $i \neq 1$}

We prove this for $i=2.$ Note that condition (A3) implies $\frac{p-q}{\sqrt{p}}=\tilde{\Omega}(k/\sqrt{n}),$ or $\alpha n (p-q)=\tilde{\Omega}(\sqrt{n p}).$ Since $p-q=\Theta\left(\frac{\log n}{n^\epsilon}\right)$ we have in equation \eqref{eq:fbart_gap} $\alpha n (p-q)>1,$ therefore the gap $\bar{f}_i^i(t)-\bar{f}_j^i(t)$ is of the same order of $\bar{f}_i^i(t),\bar{f}_j^i(t).$ For $r = \frac{2\log (n^\epsilon /L) }{\log np}$ the expected weights are given by,
\begin{eqnarray*}
E[w_l|l \in V_1] &=& L p f_1^1(r) + Lq f_2^1(r) + \sum_{j \neq 1,2} Lq f_j^1(r) \\ 
E[w_l | l \in V_2] &=& L p f_1^2(r) + Lq f_2^2(r) + \sum_{j \neq 1,2} Lq f_j^2(r)
\end{eqnarray*}  
Subtracting the above equations,
\begin{eqnarray}
E[w_l|l \in V_1]-E[w_l | l \in V_2] 
&\stackrel{(a)}{=}& L p f_1^1(r) + Lq f_2^1(r) -  L p f_1^2(r) - Lq f_2^2(r) \nonumber \\
&=& L p \bar{f}_1^1(r) + Lq \bar{f}_2^1(r) -  L p \bar{f}_1^2(r) - Lq \bar{f}_2^2(r) \nonumber \\
&& - o(Lp(\bar{f}_1^1(r)+\bar{f}_2^2(r))) \nonumber \\
&\stackrel{(b)}{\geq}& L(p-q)(\bar{f}_1^1(r)- \bar{f}_2^1(r)) - o(Lp\bar{f}_1^1(r)) \nonumber \\
&\stackrel{(c)}{>}& 0 \nonumber 
\end{eqnarray}
Steps (a), (b) follow from symmetry since $\bar{f}_1^1(r)=\bar{f}_2^2(r),$ and $\bar{f}_i^j(r)$ are all equal for $i \neq j.$ Step (c) uses Claim 1. Hence the proof.

{\bf Claim 3:} {\em $w_i$ satisfy condition (A2)}

Again for $r = \frac{2\log (n^\epsilon /L) }{\log np},$ and $p-q=\Theta\left(\frac{\log n}{n^{\epsilon}} \right)$ we have $f_1^1(r)-f_2^1(r) = \Theta(n^\epsilon/L).$ Then,
\begin{eqnarray*}
\sigma_1(R)-\sigma_2(R) &=& E[w_i|i \in V_1]-E[w_i|i \in V_2] \\
&=& \Theta(L(p-q)n^\epsilon/L) = \Theta(\log n) 
\end{eqnarray*}
Also using Chernoff bound with high probability $\gamma_2 = O(\sqrt{\log n}).$ Now for $p=\Theta\left(\frac{\log n}{n^\epsilon}\right),q=\Theta\left(\frac{\log n}{n^\epsilon}\right),$ $p-q=\Theta\left(\frac{\log n}{n^\epsilon}\right),$ under condition (A3) with $k=\Theta\left( n^{(1-\epsilon)/2}\right),$ (A2) requires $\gamma_2$ to satisfy the condition

\begin{eqnarray*}
\gamma_2&=&O\left((\sigma_1(R)-\sigma_2(R))\min\left\{\frac{1}{\xi(n)},\frac{\sqrt{\log n}}{\xi(n)}-1\right\}\right) \\
&=&O\left(\min\left\{\frac{\log n}{\xi(n)},\frac{(\log n)^{1.5}}{\xi(n)}-\log n\right\}\right)
\end{eqnarray*}

This is satisfied since $\xi(n)=o(\sqrt{\log n})$ and $\gamma_2=O(\sqrt{\log n}).$ Hence condition (A2) holds with high probability. 
\end{proof}



\end{document}